\documentclass[aps,prb,showpacs,twocolumn]{revtex4}
\usepackage{amsmath}
\usepackage{mathpazo,epsfig,amssymb,amsmath,color,graphicx,rotating,subfloat,booktabs}
\usepackage[version=3]{mhchem}
\usepackage{amsfonts}
\usepackage{amsmath}
\usepackage{amssymb}
\usepackage{graphicx}
\usepackage{subfigure}
\usepackage{subfloat}

\input{tcilatex}
\begin{document}

\title{Approximate Theory of Temperature Coefficient of Resistivity of
Amorphous Semiconductors}
\author{Ming-Liang Zhang and D. A. Drabold}
\affiliation{Department of Physics and Astronomy, Ohio University, Athens, Ohio 45701}

\begin{abstract}
In this paper, we develop an approximate theory of the temperature
coefficient of resistivity (TCR) and conductivity based upon the recently
proposed Microscopic Response Method. By introducing suitable approximations
for the lattice dynamics,  localized and extended
electronic states, we produce new explicit forms for the conductivity and
TCR, which depend on easily accessible material parameters. The theory is in
reasonable agreement with experiments on a-Si:H and a-Ge:H. A long-standing
puzzle, a \textquotedblleft kink\textquotedblright\ in the experimental $%
\log _{10}\sigma $ vs. 1/T curve, is predicted by the theory and attributed
to localized to extended transitions, which have not been properly handled
in earlier theories.
\end{abstract}

\pacs{71.23.An, ~71.38.Fp, ~71.38.Ht.  }
\keywords{eigenvector of normal modes, conductivity, atomic displacement}
\maketitle


\section{Introduction}

\label{intro}

The temperature coefficient of resistivity (TCR) of an amorphous
semiconductor (AS) is not only an important quantity in transport theory,
but also a critical parameter controlling the sensitivity of uncooled
microbolometers employed in thermal imaging \textquotedblleft night
vision\textquotedblright\ applications\cite{motda,str}.

The conventional approach to transport coefficients is the kinetic method
(Boltzmann or master equations etc.). However, this is not applicable even
to crystalline semi-metals and semiconductors (the so-called Landau-Peierls
criterion)\cite{pei,pei5,v10}. Comparing to metals, the low carrier
concentration in these materials results in a lower kinetic energy of
carriers. Thus neither the elastic scattering by disorder, nor the inelastic
scattering by a phonon has a well-defined transition probability per unit
time\cite{pei,pei5,v10}. In AS, the strong electron-phonon interaction of
localized states requires a reorganization of the vibrational configuration
for any transition involving localized state(s)\cite{epjb,pss}. For these
intrinsic multi-phonon transitions, the energy conservation between initial
and final electronic states (a basic condition of Fermi's golden rule) \cite%
{pei,pei5,v10}, is violated more seriously than that for single-phonon
emission and absorption.

In addition, transitions between localized and extended states (LE and EL)
are not treated adequately in a kinetic approach. The Miller-Abrahams theory%
\cite{ma} and its extensions suppose that LE and EL transitions do not
directly contribute to conduction, and only maintain the distribution of
carriers between localized states and extended states in thermal equilibrium
(when an external electric field is absent)\ or in the non-equilibrium
stationary state (when an external field is present). Electrical conduction
is fulfilled by the transition from a localized state to another localized
state (LL) and the transition from an extended state to another extended
state (EE)\cite{str,motda,over}. The theory of phonon-induced delocalization
and the theory of transient current excited by photon have heuristically
estimated conductivity from LE and El transitions.

Rigorous expressions for the conductivity and Hall mobility in AS have been
obtained in the microscopic response method (MRM)\cite{short,pss}. These
expressions require transition amplitudes rather than transition probability
per unit time\cite{kubo}. Thus the long-time limit required in a kinetic
approach\cite{pei,pei5} is avoided. To the lowest order self-consistent
approximation, there are 29 processes contributing to conductivity and 10
processes contributing to Hall mobility\cite{pss}. For example, in a n-doped
AS the conductivity from LE transitions driven solely by an external field is%
\cite{pss}%
\begin{equation*}
\left\{
\begin{array}{c}
\func{Re} \\
\func{Im}%
\end{array}%
\right. \sigma _{\alpha \beta }(\omega )=-\frac{N_{e}e^{2}}{2\Omega _{%
\mathbf{s}}}\sum_{AB}\func{Im}\frac{(w_{AB}^{\alpha }-v_{BA}^{\alpha
})(v_{BA}^{\beta })^{\ast }}{(E_{A}^{0}-E_{B}^{0})}
\end{equation*}%
\begin{equation}
i[I_{BA+}\pm I_{BA-}][1-f(E_{B}^{0})]f(E_{A}^{0}),\text{ \ \ }\alpha ,\beta
=x,y,z,  \label{lec}
\end{equation}%
where the real part takes the upper sign, imaginary part the lower sign. $%
\Omega _{\mathbf{s}}$ is the physical infinitesimal volume element used to
take spatial average. An AS can be viewed as uniform when we measure its
properties (e.g. conductivity) at a linear length scale larger than\cite{com}
10nm. If we take $\Omega _{\mathbf{s}}$ as a sphere with a radius larger
than 5nm, then the choice of the center $\mathbf{s}$ of $\Omega _{\mathbf{s}%
} $ inside the AS will not affect\cite{short,pss} $\sigma _{\alpha \beta }$.
$N_{e}$ is the number of carriers in the conduction band inside $\Omega _{%
\mathbf{s}}$, and $f$ is the Fermi distribution function. The velocity
matrix elements in Eq.(\ref{lec}) are defined by

\begin{equation}
v_{BA}^{\alpha }=-\frac{i\hbar }{m}\int d^{3}x\chi _{B}^{\ast }(\mathbf{r})%
\frac{\partial }{\partial x_{\alpha }}\phi _{A}(\mathbf{r}-\mathbf{R}_{A}),
\label{vle}
\end{equation}%
and%
\begin{equation}
w_{AB}^{\alpha }=-\frac{i\hbar }{m}\int d^{3}x\phi _{A}(\mathbf{r}-\mathbf{R}%
_{A})\frac{\partial }{\partial x_{\alpha }}\chi _{B}^{\ast }(\mathbf{r}),
\label{wle}
\end{equation}%
where $E_{A}^{0}$ and $\phi _{A}$ are the eigenvalue and eigenfunction of
localized state A. We will use letter $A$ with or without a natural number
subscript to denote a localized state, similarly $E_{B}^{0}$ and $\chi _{B}$
are the eigenvalue and eigenfunction of extended state B. $I_{B_{1}A\pm }$
arise from integrating out the vibrational degrees of freedom, and are
functions of external field frequency $\omega $:%
\begin{equation*}
I_{B_{1}A\pm }(\omega )=\exp \{-\frac{1}{2}\sum_{\alpha }\coth \frac{\beta
\hbar \omega _{\alpha }}{2}(\theta _{\alpha }^{A})^{2}\}
\end{equation*}%
\begin{equation*}
\int_{-\infty }^{0}dse^{is(\pm \omega +\omega _{AB_{1}})}
\end{equation*}%
\begin{equation}
\exp \{\frac{1}{2}\sum_{\alpha }(\theta _{\alpha }^{A})^{2}[\coth \frac{%
\beta \hbar \omega _{\alpha }}{2}\cos s\omega _{\alpha }-i\sin s\omega
_{\alpha }]\},  \label{tij}
\end{equation}%
where $\omega _{AB}=(E_{A}^{0}-E_{B}^{0})/\hslash $, $\omega _{\alpha }$ is
the frequency of the $\alpha ^{\text{th}}$ ($\alpha =1,2,\cdots 3\mathcal{N}$%
) normal mode, $\mathcal{N}$ is number of atoms inside $\Omega _{\mathbf{s}}$%
. Denote $\Theta _{\alpha }^{A}$ as the shift in the origin of the $\alpha ^{%
\text{th}}$ mode induced by the electron-phonon (e-ph) interaction in a
localized state\cite{epjb,pss} A, $\theta _{\alpha }^{A}=\Theta _{\alpha
}^{A}(M_{\alpha }\omega _{\alpha }/\hbar )^{1/2}$. To make the narration
specific, we hereafter discuss conduction band transport only. For transport
processes in the valence band, one may repeat the discussion \textit{mutatis
mutandis}.

To calculate conductivity strictly, one needs (i) the eigenvalues and
eigenvectors of single-electron states and (ii) the eigenfrequencies and
eigenvectors of the normal modes and the electron-phonon coupling. These can
be approximately obtained by one step of \textit{ab initio} molecular
dynamics for an optimized configuration. Then one can compute (i) $%
v_{BA}^{\alpha }$ for all localized states and extended states; (ii) $\theta
_{\alpha }^{A}$ for all normal modes in each localized states; (iii) time
integrals $I_{B_{1}A\pm }$ for a given $\omega $; and (iv) sum over all
localized states and extended states $\sum_{AB}$. Although the result
obtained in this way should be accurate and predictive, it is useful to
develop an approximate theory, which also provides functional dependence of
transport on various material parameters.

In this paper, we will first present a tractable model for the conductivity
and Hall mobility in AS. Then we will use this model to simplify the
conductivity expressions obtained in the MRM for the three simplest
transitions: LL, LE and EL transitions driven solely by external field, cf.
Fig. 2a, 2b and 6a of [\onlinecite{pss}]. The conductivity from EE
transition caused by disorder has been solved in the coherent potential
approximation\cite{but,ban}, exhibits weak temperature dependence, and we
will not consider it further. 

The outline of the paper is as following. In Sec.\ref{asum} we describe our
approximation for the lattice vibrations and e-ph in coupling.
In Sec.\ref{le}, we first illustrate that the MRM\ conductivity can be put
in the customary\ form of relaxation time approximation and of Greenwood
formula. At moderately high temperature, we invoke an asymptotic expansion
to simplify the time integrals $I_{B_{1}A\pm }$. Under the approximations
introduced in Sec.\ref{asum}, one can (i) obtain the velocity matrix
elements analytically; (ii) partially carry out the two-fold summations over
the initial and final electronic states. The conductivity from EL
transitions is obtained in Sec.\ref{el}. The conductivity from LL
transitions is calculated in Sec.\ref{ll}. The matrix elements of electronic
velocity could be carried out in a spherical coordinate system analytically.
The conductivity from the LE transitions is the same order of magnitude as
those from the LL transitions. Below a crossover temperature T$^{\ast }$,
the later is larger; above T$^{\ast }$, the former is larger. This
phenomenon is the main reason for the kink in the experimental $\log
_{10}\sigma $ vs. 1/T curve. As a demonstration, the numerical results for
n-doped a-Si:H and a-Ge:H samples are given.

\section{Approximate implementation of MRM}

\label{asum}

\subsection{Vibrations}

\label{dm}

To calculate the\ e-ph interaction for a localized state, we need the
transformation matrix between the atomic displacements and normal modes\cite%
{epjb}. Because most amorphous materials are isotropic\cite{str,motda} and
only acoustic modes are important for the e-ph interaction in one component
semiconductors\cite{han}, one can use the acoustic dispersion relation for
the vibrational spectrum:%
\begin{equation}
\omega _{\mathbf{k}}=\overline{c}k,\text{ \ }k=|\mathbf{k}|  \label{pers}
\end{equation}%
where $\omega _{\mathbf{k}}$ is the angular frequency for any mode
characterized by wave vector $\mathbf{k}$. For every $\mathbf{k}$, there are
one longitudinal and two transverse modes. We will use $\mathbf{k}\tau $ to
label a normal mode, where $\tau =1,2,3$ is the index of phonon branches\cite%
{v7}. Although translational invariance is destroyed in AS, standing wave
modes are still well-defined. Here, $\overline{c}$ is the average speed of
sound:%
\begin{equation}
\frac{3}{\overline{c}^{3}}=\frac{2}{c_{t}^{3}}+\frac{1}{c_{l}^{3}},
\label{sv}
\end{equation}%
where $c_{t}$ and $c_{l}$ are the speeds of transverse and longitudinal
waves which are determined by\cite{v7} the bulk modulus $B$ and shear
modulus $\mu $. The cutoff wave vector $k_{D}=(6\pi ^{2}n_{a})^{1/3}$ is
determined by the number density $n_{a}=\mathcal{N}/V$ of atoms, where $V$
is the volume of an AS, $\mathcal{N}$ is total number of atoms\cite{ash}. $%
n_{a}$ can be inferred from the observed mass density $\rho _{m}$. For a-Si
and a-Ge, $\rho _{m}$, $B$, $\mu $\cite{over,com}, $k_{D}$ and $\overline{c}$
are listed in Table \ref{pav}.
\begin{table}[tbp]
\caption{Parameters for vibrational spectrum}
\label{pav}%
\begin{tabular}{llllll}
\hline\hline
& B(GPa) & $\mu $(GPa) & $\overline{c}$($10^{3}$m/s) & $k_{D}$(\AA $^{-1}$)
& $\rho _{m}$(g/cm$^{3}$) \\ \hline
a-Si\cite{com,bey,dwi} & 100 & 52 & 6.21 & 1.44 & 2.33 \\
a-Ge\cite{com,bey,dwi} & 75 & 41 & 3.08 & 1.38 & 5.33 \\ \hline
\end{tabular}%
\end{table}
For a-Si, the Debye frequency $\omega _{D}$ is $8.91\times 10^{13}$Hz, not
far from the observed cut-off frequency\cite{kam} $70$meV$=1.07\times
10^{14} $Hz.

It is convenient to use $\{x_{3(j-1)+1},x_{3(j-1)+2},x_{3(j-1)+3}\}$ to
represent the vibrational displacement vector $\mathbf{u}_{j}=%
\{u_{jx},u_{jy},u_{jz}\}$ for the $j^{th}$ atom ($j=1,2,3\cdots \mathcal{N}$%
). Denote $\Theta _{\alpha }(\alpha =1,2,\cdots ,3\mathcal{N})$ as the
normal coordinate of the $\alpha ^{th}$ mode, so that the atomic
displacements and the normal modes are related by
\begin{equation}
x_{m}=\sum_{\alpha }\Delta _{m\alpha }\Theta _{\alpha },\text{ }m=1,2,\cdots
,3\mathcal{N}  \label{d2n}
\end{equation}%
where $\Delta $ is the minor of the determinant $|\Lambda _{jl}-\omega
^{2}M_{j}\delta _{jl}|$ ($j,l=1,2,3\cdots 3\mathcal{N}$), $\Lambda $ is the
force constant matrix\cite{v1}. When we use $\mathbf{k}\tau $ to label
modes, $\sum_{\alpha }\rightarrow \sum_{\mathbf{k}\tau }$.

For a localized state, the shifts in the origins of normal modes caused by
the e-ph interaction are the key quantities to determine the reorganization
energy for transitions involving the localized state\cite{epjb}. The shift
in origin is determined\cite{epjb} by $\Lambda ^{-1}$, $\Delta $ and the
e-ph coupling constant. $\Lambda ^{-1}$ and $\Delta $ are complicated for a
system with many atoms. To avoid using $\Lambda ^{-1}$ and find a more
practical $\Delta $, we use a continuum to model the discrete random network
of AS. In a continuum one can classify the atomic vibrations according to
possible standing wave modes. There is no reciprocal lattice for AS.
Because a continuum is isotropic and has continuous translational symmetry,
the wave vectors of the possible standing waves ($\mathbf{k}$ points) is
uniformly distributed in the wave vector space (Debye sphere $S_{D}$). The $%
\mathcal{N}$ $\mathbf{k}$-points inside $S_{D}$ correspond to $3\mathcal{N}$
vibrational modes.

The atomic displacement $\mathbf{u}$ at position $\mathbf{R}$ and time $t$
satisfies the wave equation%
\begin{equation}
\frac{1}{\overline{c}^{2}}\frac{\partial ^{2}\mathbf{u}(\mathbf{R},t)}{%
\partial t^{2}}=\nabla ^{2}\mathbf{u}(\mathbf{R},t).  \label{we}
\end{equation}
The plane wave solution of Eq.(\ref{we}) is\cite{ash}
\begin{equation}
\mathbf{u}(\mathbf{R},t)=\frac{1}{\mathcal{N}^{1/2}}\sum_{\mathbf{k}\tau
}e^{i\mathbf{k}\cdot \mathbf{R}}\mathbf{e}_{\mathbf{k}\tau }\Theta _{\mathbf{%
k}\tau }e^{-i(t\overline{c}k+\varphi _{\mathbf{k}\tau })},  \label{yi}
\end{equation}%
where $\mathbf{e}_{\mathbf{k}\tau }$ is the polarization vector of mode $%
\mathbf{k}\tau $. For a one-component system\cite{ash},%
\begin{equation}
\mathbf{e}_{\mathbf{k}\tau }\cdot \mathbf{e}_{\mathbf{k}\tau ^{\prime
}}^{\ast }=\delta _{\tau \tau ^{\prime }}.  \label{jao}
\end{equation}%
$\Theta _{\mathbf{k}\tau }$ and $\varphi _{\mathbf{k}\tau }$ are the
amplitude and phase of mode $\mathbf{k}\tau $, and are determined by the
initial conditions. %
The inverse of Eq.(\ref{yi}) is%
\begin{equation}
\Theta _{\mathbf{k}\tau }e^{-i(t\overline{c}k+\varphi _{\mathbf{k}\tau })}=%
\frac{1}{\mathcal{N}^{1/2}}\sum_{\mathbf{R}}\mathbf{u}(\mathbf{R},t)\cdot
\mathbf{e}_{\mathbf{k}\tau }^{\ast }e^{-i\mathbf{k}\cdot \mathbf{R}}.
\label{ni}
\end{equation}%
The normal coordinate of mode $\mathbf{k}\tau $ is $\Theta _{\mathbf{k}\tau
}e^{-i(t\overline{c}k+\varphi _{\mathbf{k}\tau })}$, so that
\begin{equation}
\Delta _{\mathbf{u}(\mathbf{R}),\mathbf{k}\tau }=\mathcal{N}^{-1/2}e^{i%
\mathbf{k}\cdot \mathbf{R}}\mathbf{e}_{\mathbf{k}\tau },  \label{k2r}
\end{equation}%
and%
\begin{equation}
(\Delta ^{-1})_{\mathbf{k}\tau ,\mathbf{u}(\mathbf{R})}=\mathcal{N}%
^{-1/2}e^{-i\mathbf{k}\cdot \mathbf{R}}\mathbf{e}_{\mathbf{k}\tau }^{\ast }.
\label{r2k}
\end{equation}%
In other words, the $\mathbf{u}(\mathbf{R})^{\text{th}}$ column of matrix $%
\Delta ^{-1}$ is the $(\mathbf{k}\tau )^{\text{th}}$ eigenvector belongs to
the $(\mathbf{k}\tau )^{\text{th}}$ eigenvalue $(\omega _{\mathbf{k}\tau
})^{2}=(\overline{c}k)^{2}$ of the matrix of force constants. Eqs.(\ref{k2r},%
\ref{r2k}) as consequences of Eq.(\ref{we}) is contained in the Debye
assumption (\ref{pers}).

\subsection{Localized states}

\label{als} To obtain analytical expressions for the e-ph interaction in a
localized state and the velocity matrix elements, we need reasonable and
simple approximate wave functions for localized and extended states. We
assume all localized states are spherically symmetric. The difference among
localized states is expressed by the localization length\cite{motda}. For a
localized state $A$, denote $\mathbf{R}_{A}$ as the position vector of the
center, the normalized wave function is
\begin{equation}
\phi _{A}(\mathbf{r}-\mathbf{R}_{A})=\pi ^{-1/2}\xi _{A}^{-3/2}e^{-|\mathbf{r%
}-\mathbf{r}_{A}|/\xi _{A}},  \label{a15}
\end{equation}%
where $\mathbf{r}$ and $\xi _{A}$ are the coordinate of electron and
localization length\cite{nev}. Following Mott, $\xi _{A}$ is determined by
the eigenvalue $E$ of localized state $\phi _{A}$\cite{nev}:%
\begin{equation}
\xi _{E}=\frac{bZe^{2}}{4\pi \epsilon _{0}\varepsilon }(E_{c}-E)^{-1},
\label{lh}
\end{equation}%
where $Z$ is the effective nuclear charge of an atom core, $\varepsilon $ is
the static dielectric constant. $E_{c}$ is the mobility edge, $b$ is a
dimensionless constant. $b$ is determined by the shortest possible
localization length $\xi _{\min }$ with $E=0$. Realistic calculations of
tail states are given in [\onlinecite{yue,yim,ying,lud,nam}].
\begin{table}[tbp]
\caption{Parameters for electronic state}
\label{pae}%
\begin{tabular}{llllllll}
\hline\hline
& E$_{c}$(eV) & U(meV) & n$_{\text{loc}}$(\AA $^{-3}$) & Z & $\varepsilon $
& q$_{\text{TF}}$(\AA $^{-1}$) & b \\ \hline
a-Si & 0.5[\onlinecite{jj}] & 50[\onlinecite{weh}] & 5/10.86$^{3}$[%
\onlinecite{ting}] & 4 & 11.68 & 1.7 & 0.121 \\
a-Ge & 0.5 & 51 & 5/11.32$^{3}$ & 4 & 16 & 1.7 & 0.170 \\ \hline
\end{tabular}%
\end{table}

\begin{figure}[th]
\centering
\par
\subfigure[]{\includegraphics[scale=0.15]{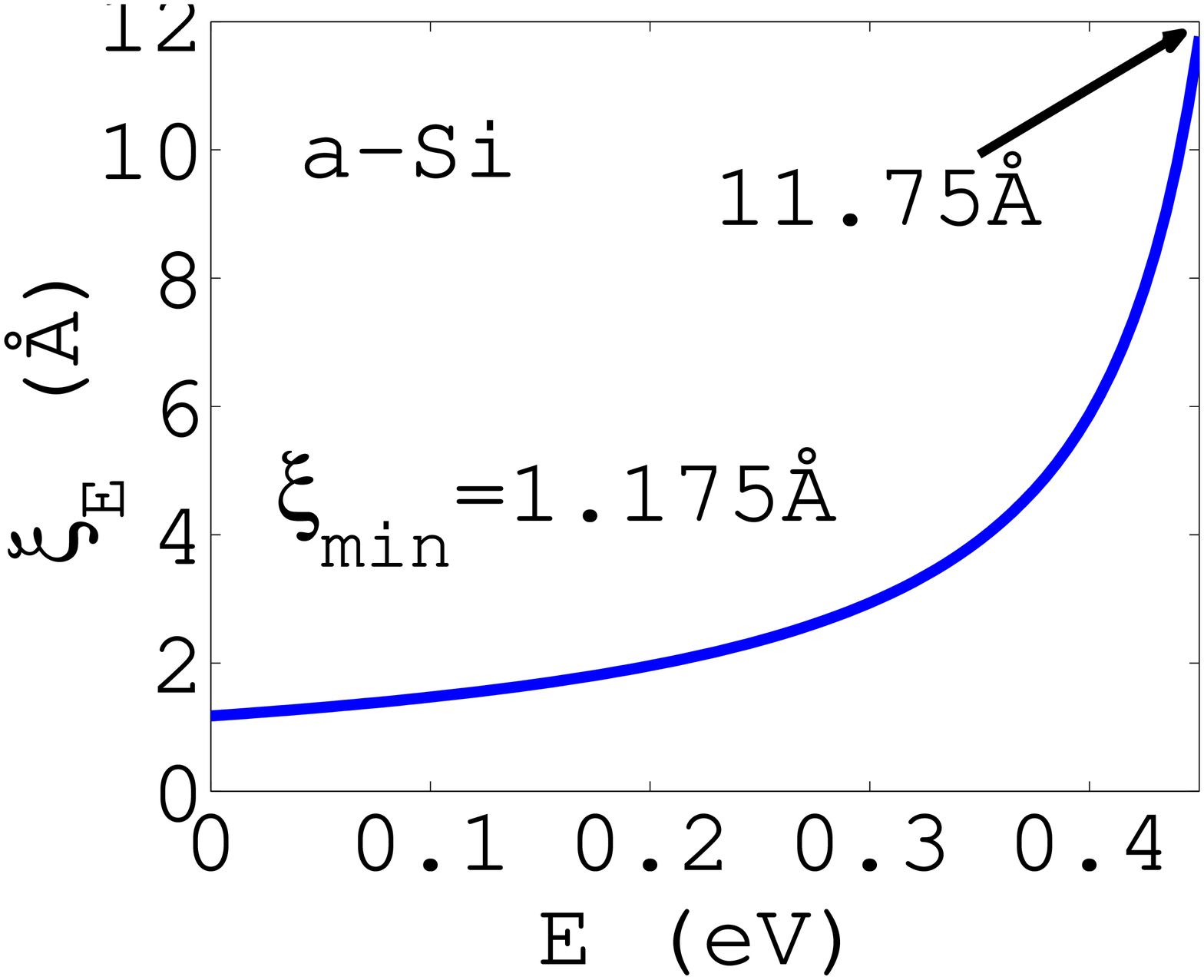}\label{locLen}}\hfill %
\subfigure[]{\includegraphics[scale=0.15]{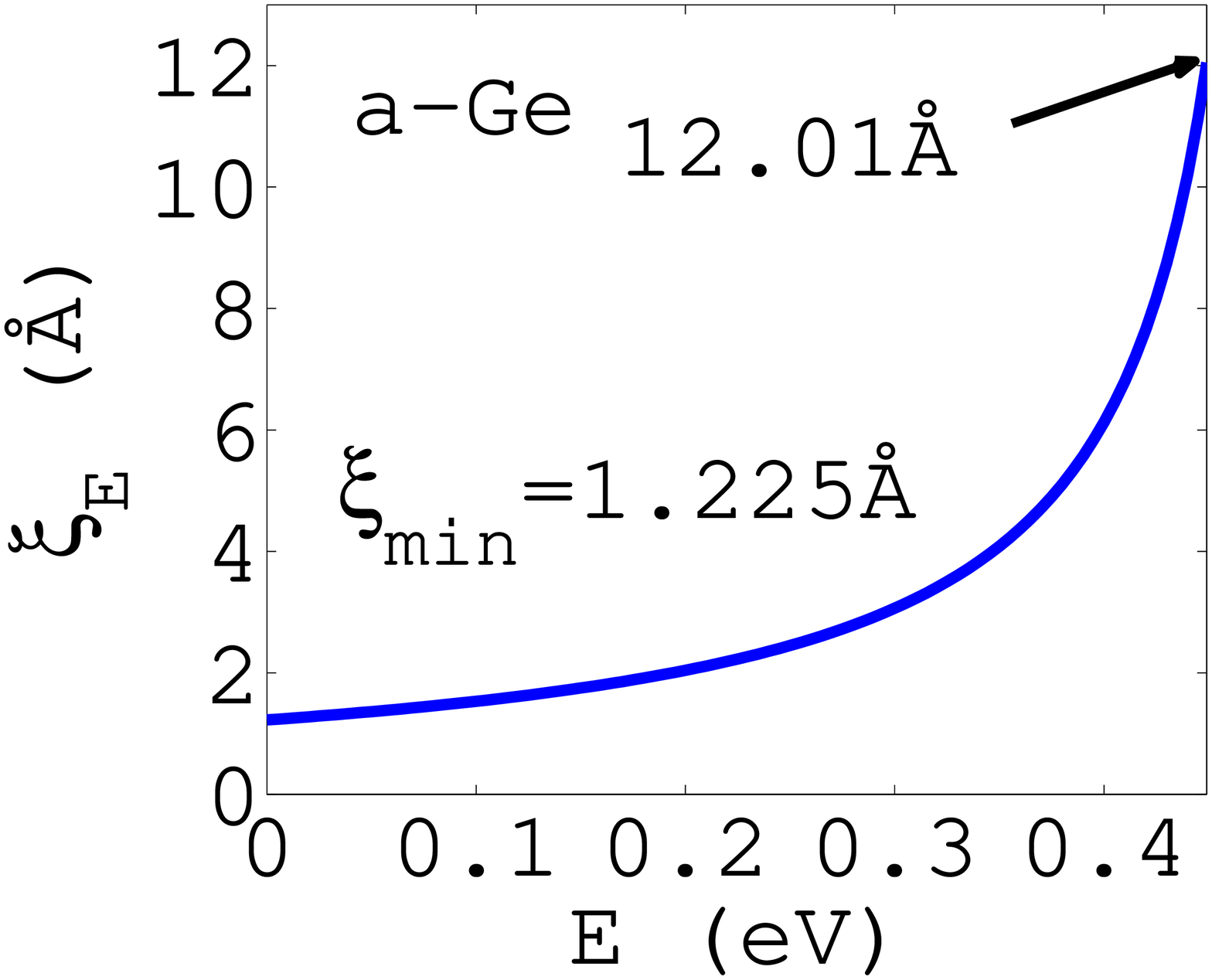}\label{GelocLen}}\hfill
\caption{Localization length as function of energy, \protect\ref{locLen}:
a-Si; \protect\ref{GelocLen}: a-Ge.}
\label{Len}
\end{figure}

The parameters\cite{str,com} for electron-core interaction and localized
state are listed in Table \ref{pae}. In a-Si:H and a-Ge:H\cite{str,com}, the
most localized states are associated with dangling bonds. The localization
length is one half the average bond length: $\xi _{\min }=2.35$\AA $/2$ and $%
2.45$\AA $/2$. Using Eq.(\ref{lh}), one has $b=0.121$ and $0.170$. The
measured value of mobility edge for a-Si is rather dispersed\cite{vis,ora}:
0.2-2eV: we will take\cite{jj} $E_{c}=0.5$eV. Fig.\ref{Len} plots
localization length vs. eigenenergy, we purposely left out a small
neighborhood $[E_{c}-U,E_{c})$ of $E$, where $U$ is the Urbach energy for
band tail. When $\xi _{E}$ is larger than the linear size of a physical
infinitesimal volume element\cite{bey} ($\thicksim $100\AA ), the
corresponding localized state acts like an extended state for purpose of
transport.

There is a distinction between a large polaron and a carrier in a weakly
localized state with $\xi $ several tens of \AA . A large polaron can move
freely before meeting a scatterer, while a localized carrier in AS is
trapped in the region where $\phi _{A}$ has support. To make a localized
carrier move, thermal activation involving a reorganization of vibrational
configuration is necessary\cite{epjb}.

Because (i) no translational invariance exists in an AS; and (ii) a
localized electronic state is confined in some finite region, the spatial
distribution of localized states needs special attention. For various
macroscopic properties, an AS can be viewed as isotropic and uniform at a
length scale larger than\cite{bey} 10nm (this effectively defines the
physical infinitesimal volume element $\Omega$). Therefore it is convenient
to describe the spatial distribution of localized states in a spherical
coordinate system. For a given origin and polar axis, the sum over localized
states $A_{1}$ can be changed into an integral over a combined spatial and
energetic distribution of localized states:
\begin{equation}
\sum_{A_{1}}\rightarrow \int_{0}^{R_{c}}R^{2}dR\int_{0}^{\pi }\sin \theta
d\theta \int_{0}^{2\pi }d\phi \int_{-\infty }^{E_{c}}dEf(R,\theta ,\phi ;E),
\label{she}
\end{equation}%
where $R$ is the distance between the origin and the center $\mathbf{R}%
_{A_{1}}$ of a localized state $\phi _{A_{1}}$, $R_{c}$ is the radius of an
AS sample, $f(R,\theta ,\phi ;E)$ is the number of localized states in a
volume element defined by ($R,R+dR),$ ($\theta ,\theta +d\theta $) and ($%
\phi ,\phi +d\phi $) with energy ($E,E+dE$), i.e. position dependent density
of states. Since a volume element with a linear size of 10nm is
representative for an AS, in the calculation of transport coefficients, one
may replace the volume $V$ of the entire AS sample with the volume $\Omega $
of a physical infinitesimal volume element. Then $R_{c}$ is the radius of $%
\Omega $.

In a physical infinitesimal volume $\Omega $, various possible atomic
configurations appear according to the proper statistical weights which
would be found in a much larger sample. Therefore the coarse-grained average
$\overline{f}$ of $f(R,\theta ,\phi ;E)$ over such a physical infinitesimal
volume element is no longer position dependent: $\overline{f}=N(E)$, where $%
N(E)$ is the usual density of states. However the weight factors in Eq.(\ref%
{she}) play an important role in determining transport properties. The
reason is that although $\overline{f}$ is independent of ($R,\theta ,\phi $%
), the transition amplitudes (velocity matrix elements) depend on the
relative position of another localized state or on the wave vector direction
of the involved extended state.

\begin{figure}[th]
\centering
\includegraphics[scale=0.2]{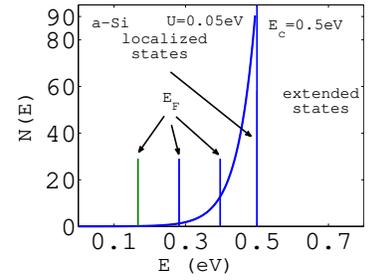}
\caption{Density of states of the conduction tail for n-doped a-Si samples:
the first three vertical lines are the Fermi energy for $n=10^{19}$, $%
10^{20} $ and $10^{21}$ cm$^{-3}$. The rightmost vertical line is the
mobility edge.}
\label{SiDOS}
\end{figure}

For many AS\cite{urbach,aljishi}, in the range of band tail, the density of
localized states satisfies%
\begin{equation}
\overline{f}(R,\theta ,\phi ;E)=N(E)=\frac{n_{loc}}{U}e^{-(E_{c}-E)/U},
\label{isu}
\end{equation}%
where $U$ is the Urbach energy, $n_{\text{loc}}$ is the number of localized
states per unit volume. The pre-exponential factor is determined from the
requirement that the integral of $N(E)$ over all localized energy spectrum
should be $n_{\text{loc}}$. In general $E_{c}$ and $U$ take different values
for the valence band and the conduction band\cite{aljishi}. Denote $n$ as
the carrier concentration, the Fermi energy $E_{F}$ of a weakly doped AS is:
\begin{equation}
E_{F}=E_{c}+U\ln (n/2n_{loc}),  \label{fer}
\end{equation}%
When $n\leq 2n_{loc}$, all occupied states are localized at T$=0$K. For
a-Si, the conduction band energy spectrum (\ref{isu}) is illustrated in Fig.%
\ref{SiDOS}.
We can see from Fig.\ref{locLen} and Eq.(\ref{isu}) that most localized
states in a-Si have a localization length in the range 6-12\AA . In
approximation (\ref{isu}), the density of states $N(E)$ of localized states
reaches its maximum at $E_{c}$. Therefore, the most probable localization
length is $\overline{\xi }=cZe^{2}/(4\pi \epsilon _{0}\varepsilon U)$. For
a-Si, $\overline{\xi }=11.75$\AA . This is consistent with various
experiments\cite{yosh,yas,iva,qg,stut,how,lou}.

Making use of relation (\ref{lh}), the integral over the energy eigenvalues
of localized states is converted into an integral over localization lengths:%
\begin{equation}
\int_{-\infty }^{E_{c}}dE_{A}N(E_{A})\rightarrow \frac{bZe^{2}n_{loc}}{4\pi
\epsilon _{0}\varepsilon U}\int_{0}^{\infty }\frac{d\xi }{\xi ^{2}}\exp (-%
\frac{bZe^{2}}{4\pi \epsilon _{0}\varepsilon U\xi }).  \label{slm}
\end{equation}%
Comparing Eq.(\ref{she}) with the sum over states $\sum_{\mathbf{k}%
}\rightarrow \int_{BZ}\frac{Vd^{3}k}{(2\pi )^{3}}$ in a crystal is helpful,
where $\mathbf{k}$ is the wave vector of a Bloch state in Brillouin zone, $V$
is volume of the crystal. The matrix elements behind $\sum_{\mathbf{k}}$ may
depend on the direction of $\mathbf{k}$, $d^{3}k=k^{2}dk\sin \theta _{%
\mathbf{k}}d\theta _{\mathbf{k}}d\phi _{\mathbf{k}}$ takes into account the
dependence on the two wave vectors of two Bloch states.

\subsection{Extended states}

\label{aes}

If one imagines that an AS is obtained from deforming its reference crystal,
an extended state in the AS can be viewed as a superposition of a principal
Bloch wave with a given wave vector and its scattered secondary waves\cite%
{vky,scm}. The scattered waves are produced by scattering the principal
Bloch wave with the disorder potential (the difference between the potential
energy in the AS and that in its reference crystal)\cite{vky,scm}. Excepting
the EE transitions driven by external field, we may approximate an extended
state $\chi _{B_{1}}(\mathbf{r})$ by a plane wave with certain momentum $%
\mathbf{p}$, and its eigenenergy is that of the plane wave:
\begin{equation}
\chi _{B_{1}}=V^{-1/2}e^{i\mathbf{p}\cdot \mathbf{r}/\hbar },\text{ \ \ }%
E_{B_{1}}=p^{2}/2m,  \label{ted}
\end{equation}%
where $V$ is the volume of AS sample, the energy zero-point of extended
states is at the mobility edge $E_{c}$. An extended state in an AS is
labeled by the wave vector of its principal Bloch wave.
The sum over extended states becomes an integral over momentum: $%
\sum_{B_{1}}\rightarrow \int \frac{Vd^{3}p}{(2\pi \hbar )^{3}}$.

\subsection{Interaction between a carrier and an atomic core}

In a solid, the attraction to an electron from an atomic core may be crudely
approximated by a screened Coulomb potential\cite{ash}
\begin{equation}
V(\mathbf{r})=\frac{Ze^{2}}{4\pi \epsilon _{0}\varepsilon }\frac{e^{-q_{TF}r}%
}{r},  \label{a14}
\end{equation}%
where $\mathbf{r}$ is the position of electron relative to an atomic core. $%
q_{TF}=2.95(r_{s}/a_{0})^{-1/2}$\AA $^{-1}$ is the Thomas-Fermi wave vector,
$r_{s}/a_{0}$ is a number about 2 to 6.
For a-Si:H\cite{str} and a-Ge\cite{com}, we take the value for c-Si and
c-Ge: $q_{TF}=1.7$\AA $^{-1}$.

\subsection{Electron-phonon coupling in a localized state}

We consider the mean e-ph interaction in a localized state $\phi _{A}$. The
e-ph interaction Hamiltonian is%
\begin{equation}
H_{\text{e-ph}}=\sum_{n\sigma }u_{n\sigma }\frac{\partial V(\mathbf{r}-%
\mathbf{R}_{n})}{\partial X_{n\sigma }},\text{ }\sigma =x,y,z,  \label{ep}
\end{equation}%
where $\mathbf{R}_{n}$ ($X_{nx},X_{ny},X_{nz}$) is the position vector of
the $n^{th}$ atom, $u_{n\sigma }$ is the $\sigma ^{th}$ Cartesian component
of vibrational amplitude of the $n^{th}$ atom. Usually the average e-ph
interaction in state $\phi _{A}$ is written in a linear coupling form\cite%
{hol59}%
\begin{equation}
\int d^{3}x\phi _{A}^{\ast }(\mathbf{r}-\mathbf{R}_{A})H_{\text{e-ph}}\phi
_{A}(\mathbf{r}-\mathbf{R}_{A})=-\sum_{n\sigma }u_{n\sigma }g_{n\sigma }^{A},
\label{a12}
\end{equation}%
where $g_{n\sigma }^{A}$ is the e-ph coupling constant in state $\phi _{A}$.
Because we consider only localized state $\phi _{A}$, it is convenient to
shift the origin of coordinate to the center $\mathbf{R}_{A}$ of $\phi _{A}$%
. In Eq.(\ref{ep}), we sum over all the atoms in $V$ of an AS sample. In
addition, the factors $V(\mathbf{r}-\mathbf{R}_{n})$ and $\phi _{A}(\mathbf{r%
}-\mathbf{R}_{A})$ in the integrand of Eq.(\ref{a12}) involve two atoms,
directly integrating over coordinate is difficult (requiring ellipsoidal
coordinate system). To obtain the coupling constant $g_{n\sigma }^{A}$, we
Fourier transform $\partial V(\mathbf{r}-\mathbf{R}_{n})/\partial X_{n\sigma
}$ in the LHS of Eq.(\ref{a12}), first carry out the integral in coordinate $%
\mathbf{r}$, then execute the integral over wave vector $\mathbf{q}$. The
final result is:%
\begin{equation*}
g_{n\sigma }^{A}=\frac{16(Ze^{2}/4\pi \epsilon _{0}\varepsilon )}{\xi ^{4}}%
\frac{X_{n\sigma }}{R_{n}^{2}}\{\frac{R_{n}}{2}\frac{e^{-2R_{n}/\xi }}{%
q_{TF}^{2}-(2/\xi )^{2}}
\end{equation*}%
\begin{equation*}
+[\frac{q_{TF}e^{-q_{TF}R_{n}}-(2/\xi )e^{-2R_{n}/\xi }}{[(2/\xi
)^{2}-q_{TF}^{2}]^{2}}-\frac{\xi }{8}\frac{e^{-2R_{n}/\xi }}{%
q_{TF}^{2}-(2/\xi )^{2}}]
\end{equation*}%
\begin{equation}
+\frac{1}{R_{n}}[\frac{e^{-q_{TF}R_{n}}-e^{-2R_{n}/\xi }}{[(2/\xi
)^{2}-q_{TF}^{2}]^{2}}-\frac{\xi ^{2}}{16}\frac{e^{-2R_{n}/\xi }}{%
q_{TF}^{2}-(2/\xi )^{2}}]\},  \label{a11}
\end{equation}%
where $R_{n}=|\mathbf{R}_{n}-\mathbf{R}_{A}|$ is the distance between the $%
n^{th}$ atom to the center $\mathbf{R}_{A}$ of localized state $\phi _{A}$.
The first term decays exponentially, the second and the third term contain
additional decay factors $R_{n}^{-1}$ and $R_{n}^{-2}$ respectively. Since
we are concerned only with localized state $\phi _{A}$, hereafter we drop
the subscript $A$ on $\xi $ and $g$.

\subsection{Polaron formation}

The static displacements of atoms induced by the e-ph interaction measure
the strength of e-ph interaction and determine whether the e-ph coupling
should be treated as a perturbation or be included in the zeroth order
Hamiltonian\cite{epjb}. The static displacement of the $m^{th}$ atomic
degree of freedom caused by the e-ph interaction in localized state $\phi
_{A}$ is\cite{epjb}%
\begin{equation}
\text{\ }x_{m}^{A0}=\sum_{p}(\Lambda ^{-1})_{mp}g_{p}^{A},\text{ }%
m,p=1,2,\cdots 3\mathcal{N},  \label{stad}
\end{equation}%
where $\Lambda ^{-1}$ is the inverse of force constant matrix. The shift $%
\Theta _{\alpha }^{A}$ in origin of the $\alpha ^{th}$ ($\alpha =1,2,\cdots 3%
\mathcal{N}$) mode by the carrier localized in state $\phi _{A}$ is\cite%
{epjb}%
\begin{equation}
\Theta _{\alpha }^{A}=\sum_{m}(\Delta ^{-1})_{\alpha m}x_{m}^{A0}.
\label{shif}
\end{equation}%
This has the physical interpretation of the polaronic relaxation due to the
e-ph coupling.

If $\Lambda ^{-1}$ and $\Delta ^{-1}$ were known analytically, we could use
Eq.(\ref{stad}) to find $\{x_{m}^{A0}\}$, and then use Eq.(\ref{shif}) to
find $\{\Theta _{\alpha }^{A}\}$. The continuum model in Sec.\ref{dm} allows
us to first find the shifts in origins $\{\Theta _{\alpha }^{A}\}$ of normal
modes in a localized state. Then static displacements $\{x_{m}^{A0}\}$ can
be obtained from Eq.(\ref{d2n}). In the continuum model, the normal modes
are labeled by wave vectors $\mathbf{k}$. Substitute Eq.(\ref{stad}) into
Eq.(\ref{shif}), notice $\Delta ^{-1}\Lambda ^{-1}=W^{-1}\Delta ^{T}$, where
$(W^{-1})_{\alpha \beta }=\delta _{\alpha \beta }M_{\alpha }^{-1}\omega
_{\alpha }^{-2}$, one concludes that
\begin{equation}
\Theta _{\mathbf{k}\tau }^{A}=M_{\mathbf{k}}^{-1}\omega _{\mathbf{k}%
}^{-2}\sum_{n\sigma }g_{n\sigma }^{A}\Delta _{n\sigma ,\mathbf{k}\tau },
\label{asf}
\end{equation}%
where $n=1,2,3\cdots \mathcal{N}$ and $\sigma =x,y,z$. Substituting Eq.(\ref%
{k2r}) into Eq.(\ref{asf}) and replacing the sum by an integral over all
space, Eq.(\ref{asf}) becomes
\begin{equation}
\Theta _{\mathbf{k}\tau }^{A}=\frac{\func{Re}\sum_{\sigma }\int_{V}d^{3}Xg_{%
\mathbf{R}\sigma }^{A}e_{\mathbf{k}\tau }^{\sigma }e^{i\mathbf{k}\cdot
\mathbf{R}}}{\mathcal{N}^{1/2}M_{\mathbf{k}}\omega _{\mathbf{k}}^{2}\Omega
_{a}},  \label{sf}
\end{equation}%
where $\Omega _{a}=V/\mathcal{N}$ is the average volume occupied by one
atom. For a-Si and a-Ge\cite{com,ash}, $\Omega _{a}\thickapprox (5.43$\AA $%
)^{3}/4$ and $(5.66$\AA $)^{3}/4$. Eq.(\ref{sf}) expresses the shift $\Theta
_{\mathbf{k}}^{A}$ in the origin of normal mode $\mathbf{k}$ with the e-ph
coupling constant $g_{n\sigma }^{A}$. 
We take $\mathbf{k}$ as the polar axis (z axis) and transform to a spherical
coordinate system, because the integrand of Eq.(\ref{sf}) does not contain
azimuthal angle $\phi $, $\{g_{nx}^{A}\}$ and $\{g_{ny}^{A}\}$ do not
contribute to $\Theta _{\mathbf{k}\tau }^{A}$. Only when $\mathbf{g}_{n}^{A}$
has a component along $\mathbf{k}$, does it contribute to $\Theta _{\mathbf{k%
}\tau }^{A}$. The integrations over the $R^{-2}$ and $R^{-3}$ terms in Eq.(%
\ref{a11}) are purely imaginary, and do not contribute to $\Theta _{\mathbf{k%
}}^{A}$. The origin shift of mode $\mathbf{k}\tau $ induced by the e-ph
interaction in localized state $\phi _{A}$ is:%
\begin{equation}
\Theta _{\mathbf{k}\tau }^{A}=\frac{1}{\mathcal{N}^{1/2}Mk^{2}\overline{c}%
^{2}}\frac{2^{7}\pi Ze^{2}/(4\pi \epsilon _{0}\varepsilon \Omega _{a}\xi
^{5})}{[q_{TF}^{2}-(2/\xi )^{2}][(2/\xi )^{2}+k^{2}]^{2}}.  \label{shid}
\end{equation}%
Because we take AS to be an isotropic continuous medium, $\Theta _{\mathbf{k}%
\tau }^{A}$ depends only on the magnitude $k$. The $k^{-2}$ divergence in
Eq.(\ref{shid}) when $k\rightarrow 0$ is caused by the Debye spectrum ($%
\omega _{\mathbf{k}}=\overline{c}k$). In a Debye model, the number of modes
per unit volume per unit angular frequency interval is\cite{ash} (2$\pi ^{2}%
\overline{c}$)$^{-1}3k^{2}$ when $k<k_{D}$.
The shift is smaller for higher wave number, decays with wave vector $%
\mathbf{k}$ as $[(2/\xi )^{2}+k^{2}]^{-2}$. Because for all materials\cite%
{ash} $q_{TF}\thicksim 1.2-2.1$\AA $^{-1}$, while $\xi >2$\AA\ for localized
states caused by topological disorder\cite{nev}, the factor $%
[q_{TF}^{2}-(2/\xi )^{2}]$ in the denominator of Eqs.(\ref{a11},\ref{shid},%
\ref{disp1},\ref{ng},\ref{bi}) will not lead to a divergent result.

Eq.(\ref{shid}) exhibits two obvious features: (i) $\Theta _{\mathbf{k}%
}^{A}>0$ for every mode $\mathbf{k}$; (ii) if $\xi _{A_{1}}<\xi _{A_{2}}$,
then $\Theta _{\mathbf{k}}^{A_{1}}>\Theta _{\mathbf{k}}^{A_{2}}$ for every
mode $\mathbf{k}$. We have shown that three-state conduction processes which
are first order in residual interactions, are the same order of magnitude as
the two-states processes discussed here\cite{pss}. Also, in the lowest order
self-consistent approximation, three- and four- state processes must be
included in the Hall mobility calculation\cite{pss}. Some of the
aforementioned transport processes involve at least two localized states. To
carry out asymptotic expansion at high temperature for such processes, the
features (i) and (ii) are essential.

\begin{figure}[th]
\centering
\subfigure[]{\includegraphics[scale=0.15]{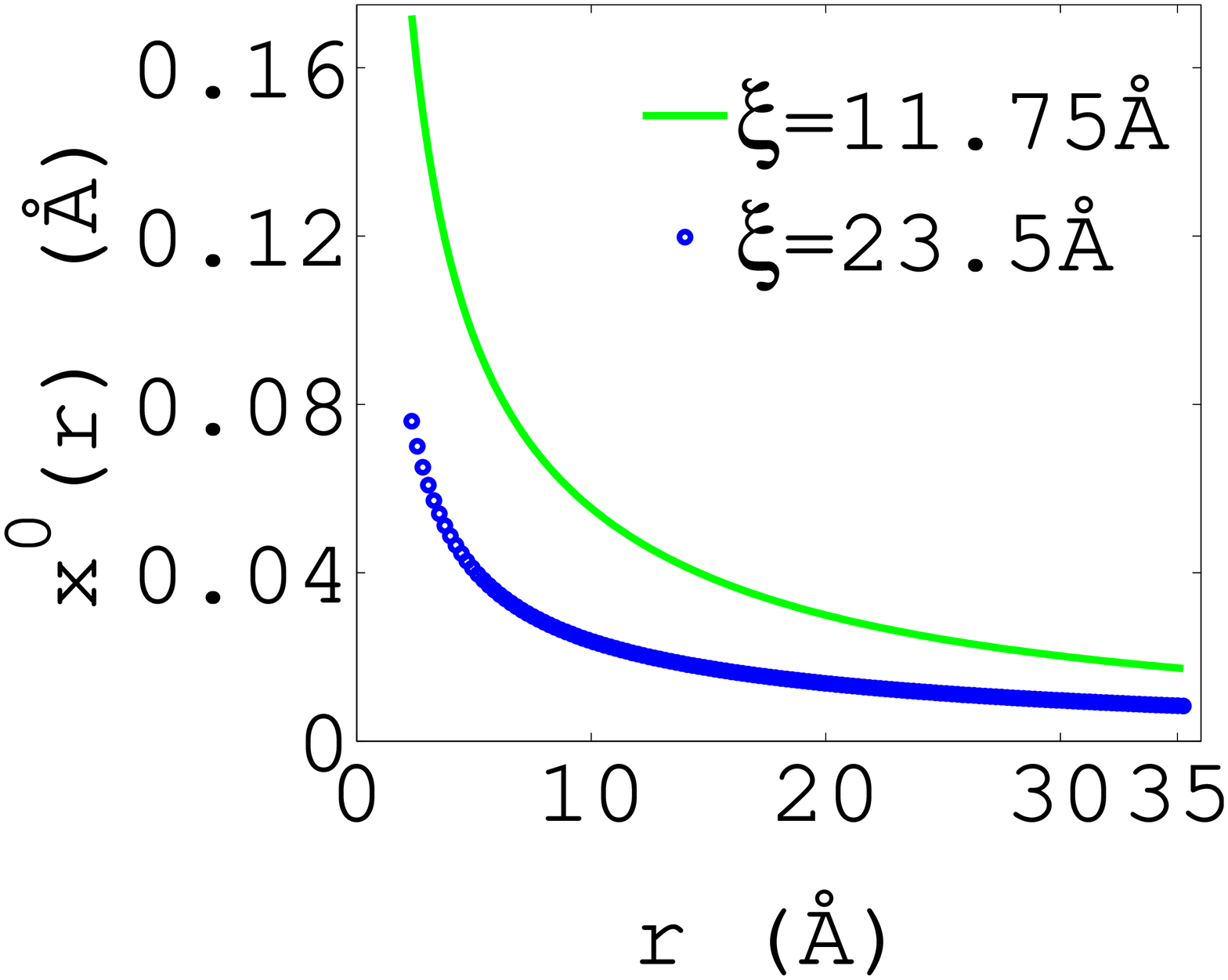}\label{disw}}\hfill %
\subfigure[]{\includegraphics[scale=0.15]{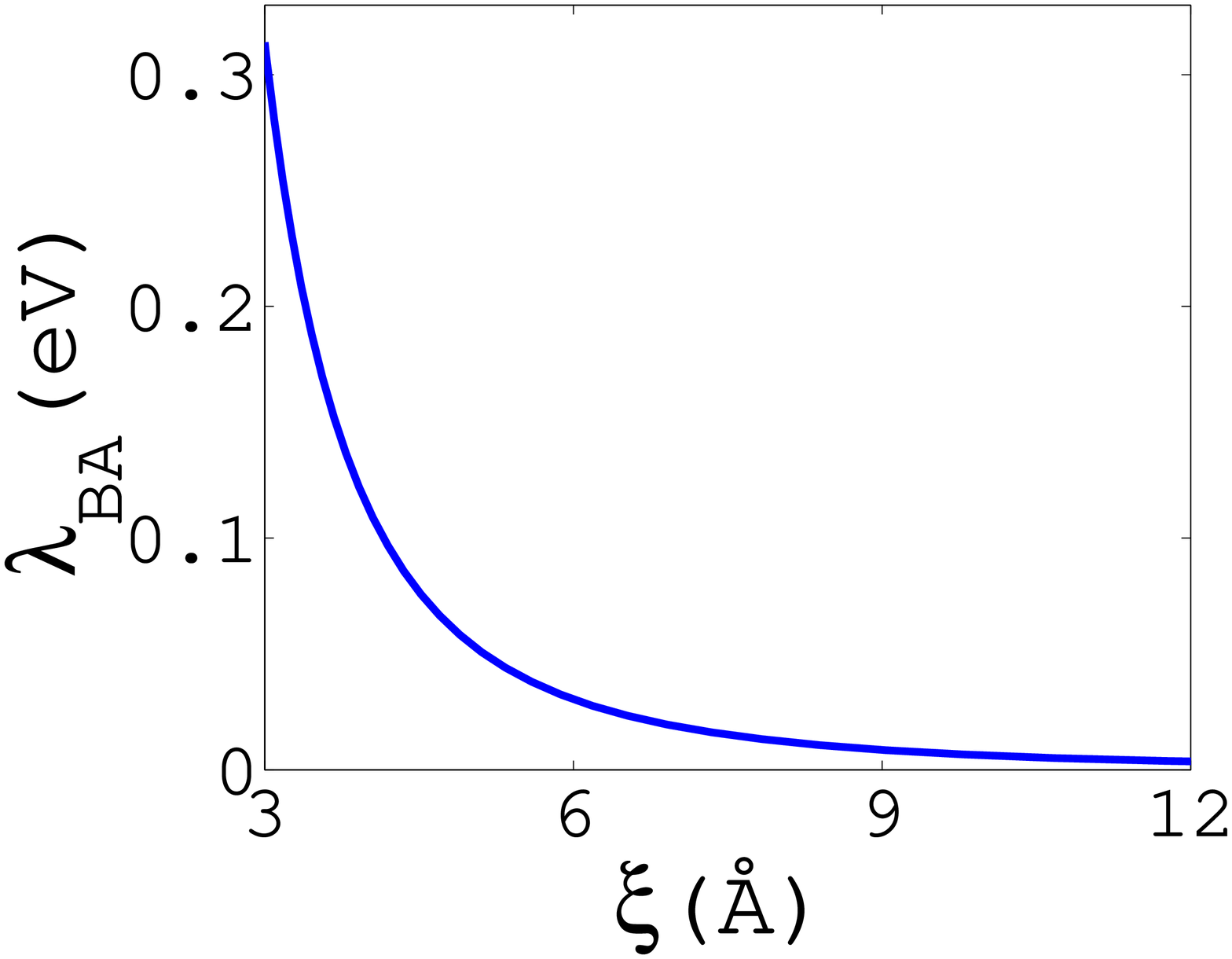}\label{binding}}\hfill
\caption{\protect\ref{disw}: Static displacements $x^{0}({\mathbf{r}})$ of
atoms in a localized state $\protect\phi_{A}$ as function of the distance $r$
to the center of $\protect\phi_{A}$ in a-Si: solid line is for $\protect\xi%
=11.75$\AA , circle line is for $\protect\xi=23.5$\AA . \protect\ref{binding}%
: The binding energy caused by e-ph interaction as function of localization
length}
\end{figure}

The static atomic displacements in localized state $A$ can be found from
Eqs.(\ref{shif},\ref{shid}):

\begin{equation}
x_{\sigma }^{0A}(\mathbf{R})=\frac{\mathcal{N}^{1/2}\Omega _{a}}{(2\pi )^{3}}%
\sum_{\tau =1}^{3}\int d^{3}ke^{i\mathbf{k}\cdot \mathbf{R}}\Theta _{\mathbf{%
k}\tau }^{A}e_{\mathbf{k}\tau }^{\sigma }.  \label{sq}
\end{equation}%
Next, substitute Eq.(\ref{shid}) into Eq.(\ref{sq}) and carry out the
integral. One finds the displacement $x^{0A}$ along the radial direction for
an atom at $\mathbf{R}$ caused by e-ph interaction in a localized state:%
\begin{equation}
x^{0A}(\mathbf{R})=\frac{4}{M\overline{c}^{2}}\frac{Z^{\ast }e^{2}/4\pi
\epsilon _{0}\varepsilon }{\xi \lbrack q_{TF}^{2}-(2/\xi )^{2}]}\frac{1-%
\frac{1}{2}e^{-2R/\xi }}{R},  \label{disp1}
\end{equation}%
where we have let $k_{D}\rightarrow \infty $ to obtain an analytic result.
It is interesting to notice that Eq.(\ref{disp1}) is similar to the wave
function of large polaron in strong coupling limit, cf. pp513-523 of [%
\onlinecite{han}].

Fig.\ref{disw} is an illustration of Eq.(\ref{disp1}) for a-Si at $\xi
=11.75 $\AA\ and $23.50$\AA\ (5 and 10 times bond length). We observe that
the more localized (smaller $\xi $) the state, the larger the atomic
displacements, i.e. the stronger e-ph interaction (larger atomic
displacements). This agrees with previous experiments and simulations\cite%
{dra,tafn}. For the hardest mode\cite{kam} $\omega =70$meV of a-Si, the
amplitude $A_{0}=$($\hbar /M\omega $)$^{1/2}$ of zero-point vibration is
0.046\AA , the amplitude $A_{th}=$($k_{B}T/M\omega ^{2}$)$^{1/2}$of thermal
vibration at 300K is 0.028\AA . Considering these two peaks of the a-Si
phonon spectrum are at\cite{kam} 20 meV ($A_{0}=0.086$\AA , $A_{th}=0.098$%
\AA ) and 60 meV ($A_{0}=0.050$\AA , $A_{th}=0.033$\AA ), the static
displacements of atoms estimated in Eq.(\ref{disp1}) are twice the amplitude
of vibrations. Comparing the root mean square of bond length fluctuation 0.2%
\AA\ (geometric disorder) from \textit{ab initio} molecular dynamics
simulation\cite{yue}, the approximate acoustic dispersion relation Eq.(\ref%
{pers}) somewhat overestimates the long wave contribution in Eqs.(\ref{sf},%
\ref{shid},\ref{sq},\ref{disp1}).

\subsection{Reorganization energy}

Unlike a carrier in an extended state, a carrier in a localized state is
confined by the disorder potential. Beyond that, the e-ph interaction
produces\cite{epjb} an additional binding energy $E_{b}^{A}$ to a localized
carrier in $\phi _{A}$:%
\begin{equation}
E_{b}^{A}=\frac{1}{2}\sum_{\alpha }M_{\alpha }\omega _{\alpha }^{2}(\Theta
_{\alpha }^{A})^{2}.  \label{bine}
\end{equation}%
Because the reorganization energy measures the energy shift from initial
vibrational configuration to the final vibrational configuration, $E_{b}^{A}$
is the same as\cite{epjb} the reorganization energy $\lambda _{BA}$ of LE
transition $\phi _{A}\rightarrow \chi _{B}$ and the reorganization energy $%
\lambda _{AB}$ of EL transition $\chi _{B}\rightarrow \phi _{A}$: $\lambda
_{AB}=\lambda _{BA}=E_{b}^{A}$. For the continuous medium model, the sum
over modes in Eq.(\ref{bine}) may be converted to an integral over the Debye
sphere in spherical coordinate system ($k,\theta ^{\prime },\phi ^{\prime }$%
):%
\begin{equation}
\lambda _{AB}=\frac{\mathcal{N}\Omega _{a}}{2(2\pi )^{3}}\sum_{\tau
=1}^{3}\int_{0}^{k_{D}}dkk^{2}\times  \label{bng}
\end{equation}%
\begin{equation*}
\int_{0}^{\pi }d\theta ^{\prime }\sin \theta ^{\prime }\int_{0}^{2\pi }d\phi
^{\prime }Mk^{2}\overline{c}^{2}(\Theta _{\mathbf{k}\tau }^{A})^{2}.
\end{equation*}%
Owing to spherical symmetry in Eq.(\ref{bng}), the direction of polar axis
is arbitrary. Substituting Eq.(\ref{shid}) into Eq.(\ref{bng}) and carrying
out the integral, one finds:
\begin{equation}
\lambda _{AB}=2\pi \frac{(2^{7}\pi Z^{\ast }e^{2}/4\pi \epsilon
_{0}\varepsilon )^{2}\xi }{2^{7}M\overline{c}^{2}\Omega _{a}[(\xi
q_{TF})^{2}-4]^{2}}  \label{ng}
\end{equation}%
\begin{equation*}
\{\frac{15}{48}\tan ^{-1}\frac{k_{D}\xi }{2}+\frac{k_{D}\xi \lbrack 1+(\frac{%
k_{D}\xi }{2})^{2}]^{-1}}{12}\times
\end{equation*}%
\begin{equation*}
([1+(\frac{k_{D}\xi }{2})^{2}]^{-2}+\frac{5}{4}[1+(\frac{k_{D}\xi }{2}%
)^{2}]^{-1}+\frac{15}{8})\}.
\end{equation*}%
Fig.\ref{binding} displays the change in binding energy with localization
length. We can see that more localized states have larger binding energy. In
other words, when a carrier leaves or enters a more localized state, the
required reorganization energy is larger, the corresponding LE and EL
transitions are more hindered.

The reorganization energy $\lambda _{A_{2}A_{1}}$ for LL transition $\phi
_{A_{1}}\rightarrow $ $\phi _{A_{2}}$ satisfies a reciprocity condition\cite%
{epjb} $\lambda _{A_{1}A_{2}}=\lambda _{A_{2}A_{1}}$, where%
\begin{equation}
\lambda _{A_{2}A_{1}}=\frac{1}{2}\sum_{\alpha }M_{\alpha }\omega _{\alpha
}^{2}(\Theta _{\alpha }^{A_{2}}-\Theta _{\alpha }^{A_{1}})^{2}\text{.}
\label{lle}
\end{equation}%
Eq.(\ref{lle}) can be expressed as%
\begin{equation}
\lambda _{A_{2}A_{1}}=|E_{b}^{A_{1}}|+|E_{b}^{A_{2}}|-B_{A_{2}A_{1}},
\label{bi1}
\end{equation}%
where $E_{b}^{A_{1}}$ is obtained from Eq.(\ref{ng}) by replacing $\xi $
with $\xi _{1}$, $\xi _{1}$ is the localization length of $\phi _{A_{1}}$. $%
B_{A_{2}A_{1}}=\sum_{\alpha }\hbar \omega _{\alpha }\theta _{\alpha
}^{A_{2}}\theta _{\alpha }^{A_{1}}$ is the interference term:%
\begin{equation}
B_{A_{2}A_{1}}=\frac{2^{7}\pi Z^{\ast }e^{2}/4\pi \epsilon _{0}\varepsilon }{%
\Omega _{a}\xi _{1}^{5}[q_{TF}^{2}-(2/\xi _{1})^{2}]}\frac{2^{7}\pi Z^{\ast
}e^{2}/4\pi \epsilon _{0}\varepsilon }{\xi _{2}^{5}[q_{TF}^{2}-(2/\xi
_{2})^{2}]}  \label{bi}
\end{equation}%
\begin{equation*}
\frac{4\pi }{M\overline{c}^{2}}\{[(2/\xi _{2})^{2}-(2/\xi _{1})^{2}]^{-2}[%
\frac{\xi _{1}^{3}}{16}\tan ^{-1}\frac{k_{D}\xi _{1}}{2}
\end{equation*}%
\begin{equation*}
+\frac{k_{D}\xi _{1}^{4}}{8(k_{D}^{2}\xi _{1}^{2}+4)}+\frac{\xi _{2}^{3}}{16}%
\tan ^{-1}\frac{k_{D}\xi _{2}}{2}+\frac{k_{D}\xi _{2}^{4}}{8(k_{D}^{2}\xi
_{2}^{2}+4)}]
\end{equation*}%
\begin{equation*}
-2[(2/\xi _{2})^{2}-(2/\xi _{1})^{2}]^{-3}[\frac{\xi _{1}}{2}\tan ^{-1}\frac{%
k_{D}\xi _{1}}{2}-\frac{\xi _{2}}{2}\tan ^{-1}\frac{k_{D}\xi _{2}}{2}]\}.
\end{equation*}%
Eqs.(\ref{ng},\ref{bi1},\ref{bi}) determined the reorganization energy $%
\lambda _{A_{2}A_{1}}$ for LL transition $\phi _{A_{1}}\rightarrow \phi
_{A_{2}}$.

\section{conductivity from LE and EL transitions driven solely by field}

\label{leel}

In this section we assemble the approximations of the proceeding section to
estimate the various contributions to the conductivity.

\subsection{LE transitions driven by field}

\label{le}

\begin{figure}[th]
\centering
\par
\subfigure[]{\includegraphics[scale=0.15]{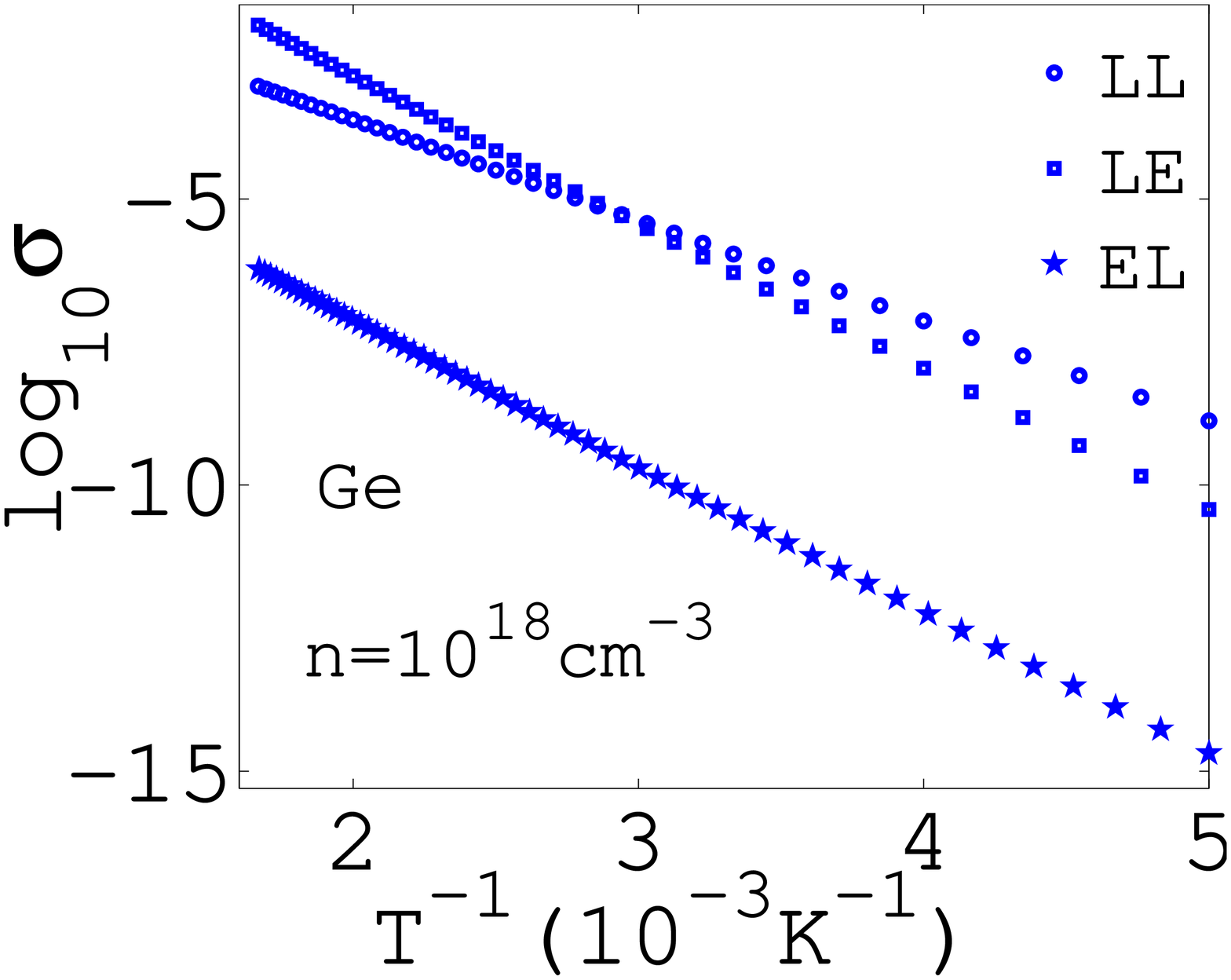}\label{18con}}\hfill %
\subfigure[]{\includegraphics[scale=0.15]{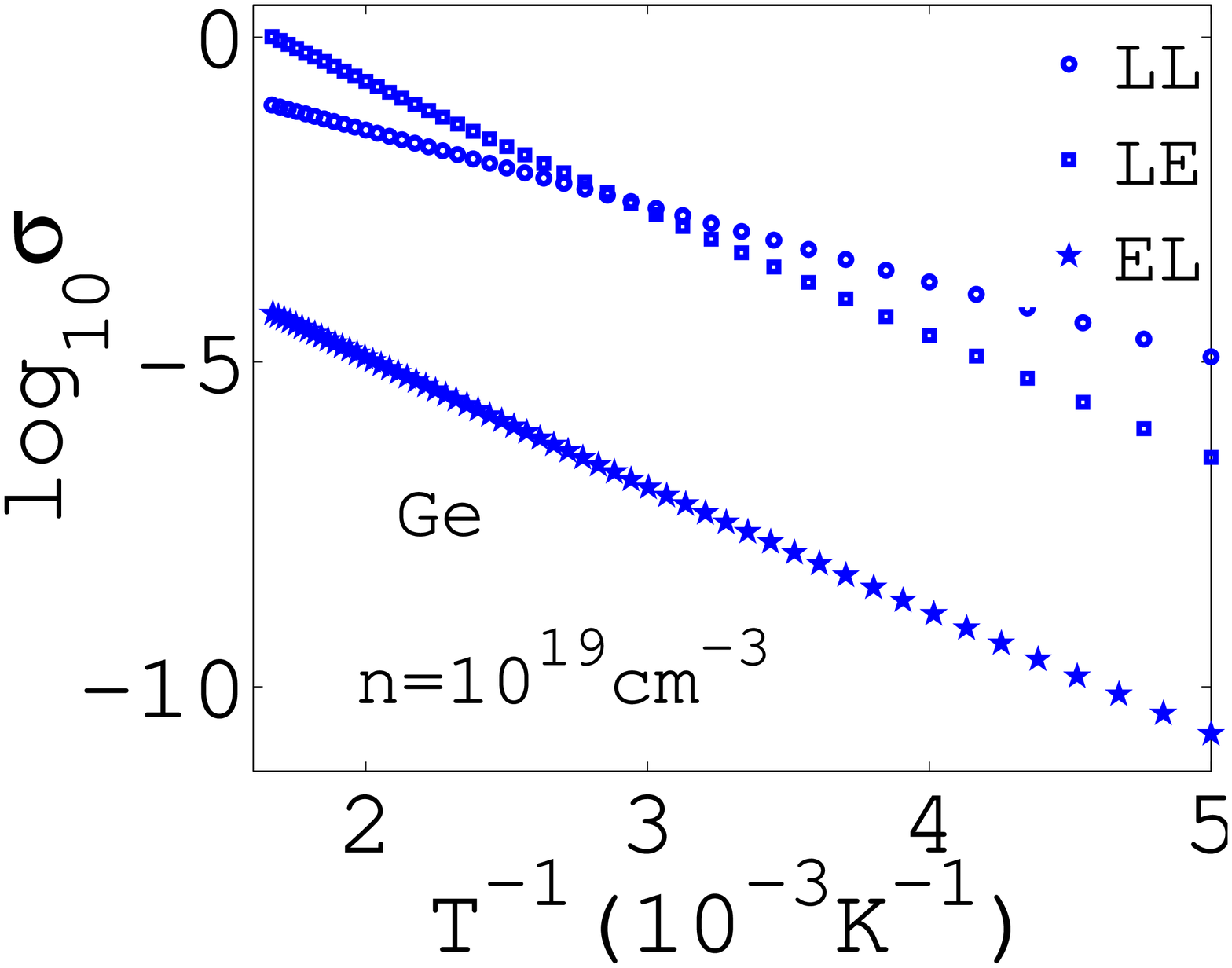}\label{19con}}\hfill %
\subfigure[]{\includegraphics[scale=0.15]{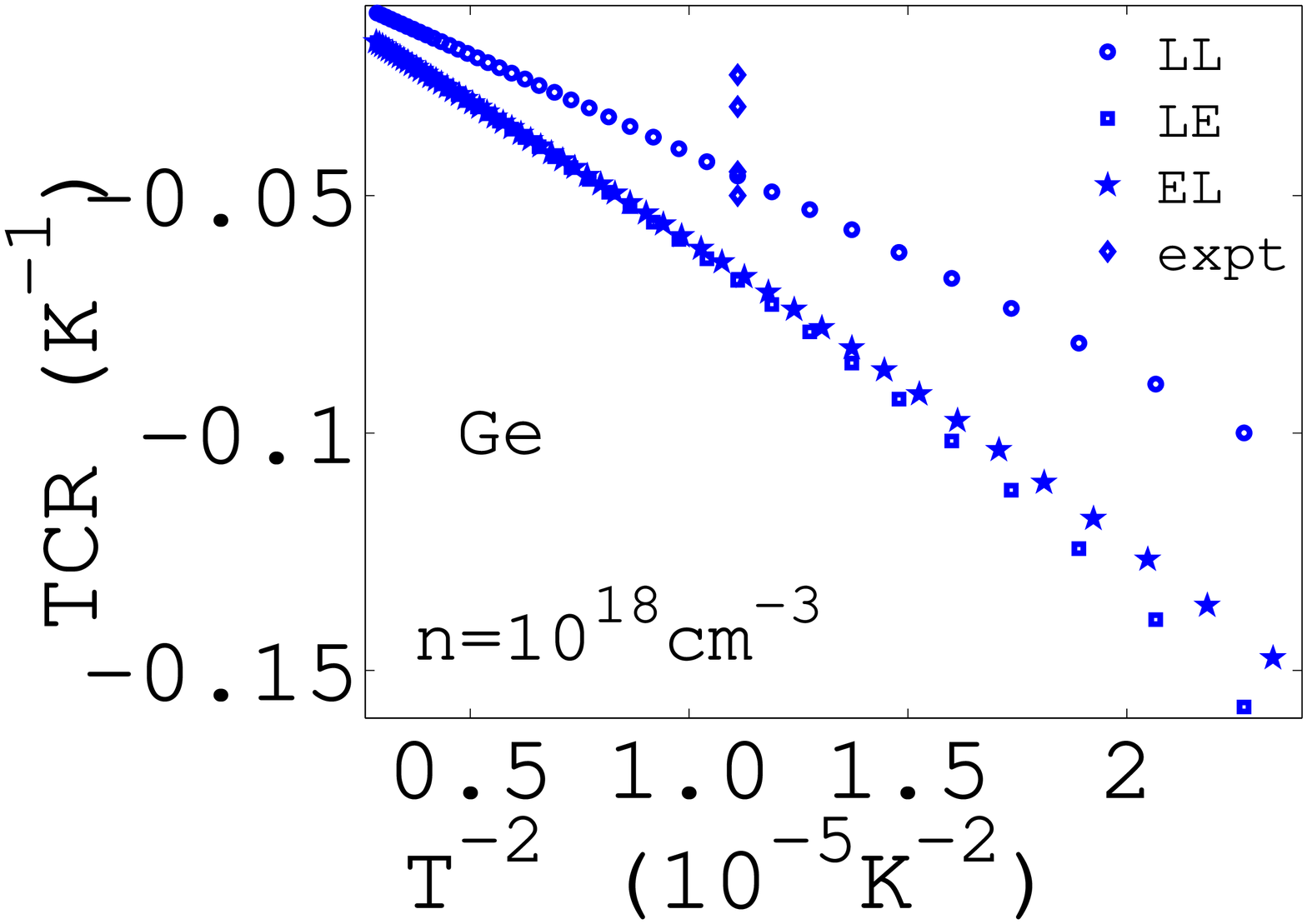}\label{18tcr}}\hfill %
\subfigure[]{\includegraphics[scale=0.15]{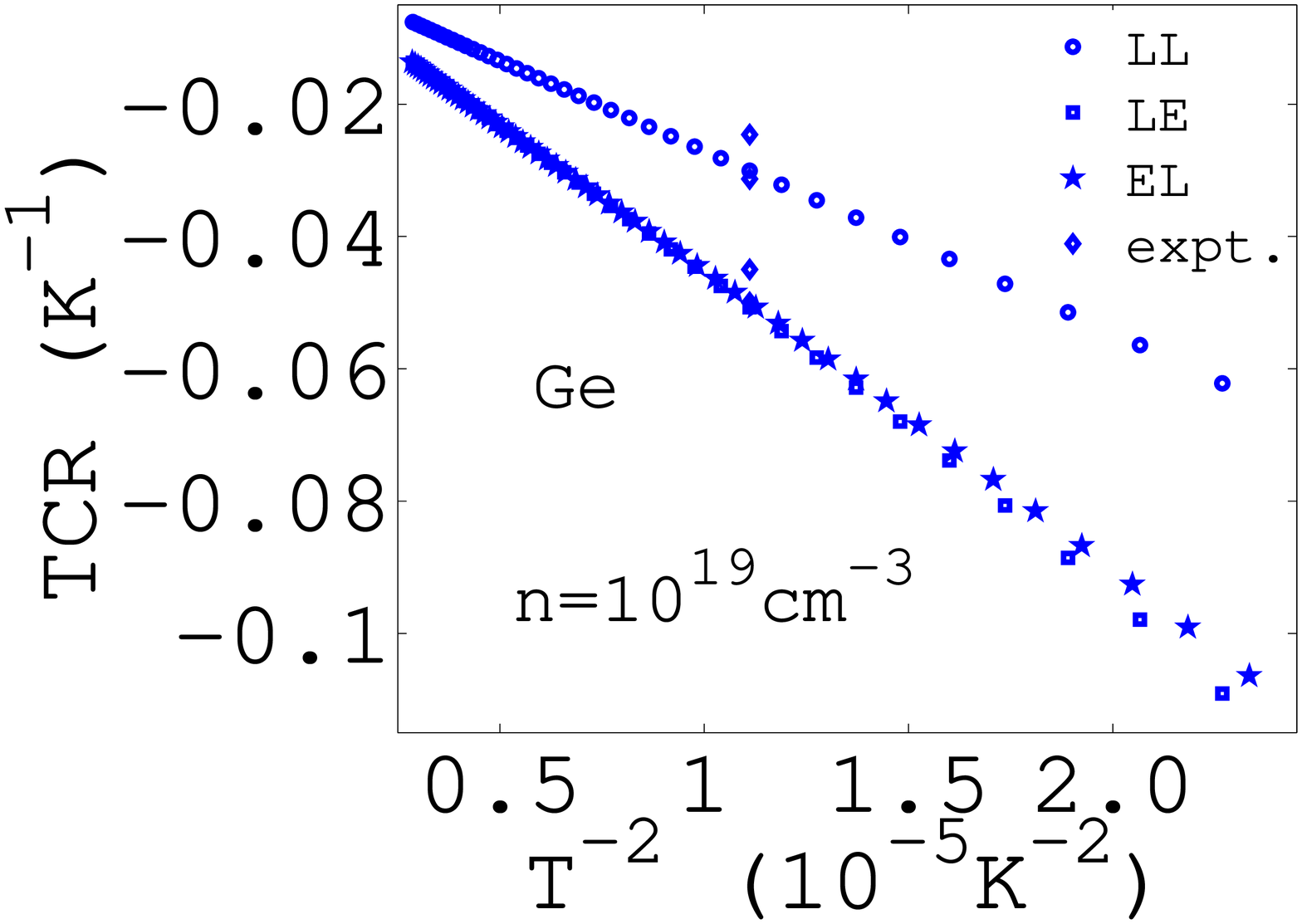}\label{19tcr}}\hfill
\caption{Conductivity and TCR as functions of temperature in two n-doped
a-Ge:H samples at $\protect\omega =0$. The experimental values are taken
from [\onlinecite{tor,ino}].}
\label{con}
\end{figure}

\subsubsection{Connection to relaxation time approximation and
Kubo-Greenwood formula}

Inside the summation of Eq.(\ref{lec}), only electronic degrees of freedom
appear. Each term can be written as:
\begin{equation}
\sigma _{\alpha \beta }^{BA}(\omega )=(m_{eff}^{BA})_{\alpha \beta
}^{-1}ne^{2}\tau ^{BA}(\omega ),  \label{fd}
\end{equation}%
where $n=N_{e}/\Omega _{s}$ is the carrier density,
\begin{equation*}
\tau ^{BA}(\omega )=\func{Im}i[I_{BA+}\pm I_{BA-}]
\end{equation*}%
may be viewed as a relaxation time, and%
\begin{equation}
(m_{eff}^{BA})_{\alpha \beta }^{-1}=-\frac{(w_{AB}^{\alpha }-v_{BA}^{\alpha
})(v_{BA}^{\beta })^{\ast }}{2(E_{A}^{0}-E_{B}^{0})},  \label{mef}
\end{equation}%
may be interpreted as the inverse of effective matrix tensor for transition $%
\phi _{A}\rightarrow \chi _{B}$. In this sense, $\sigma _{\alpha \beta
}^{B_{1}A}(\omega )$ is a generalization of the energy dependent conductivity%
\cite{motda} $\sigma _{\alpha \beta }^{E}(\omega )$. With this notation, Eq.(%
\ref{lec}) becomes%
\begin{equation}
\sigma _{\alpha \beta }(\omega )=\sum_{AB_{1}}\sigma _{\alpha \beta
}^{B_{1}A}(\omega )[1-f(E_{B_{1}})]f(E_{A}),  \label{kg}
\end{equation}%
a generalization of Kubo-Greenwood formula, Eq.(2.11) of [%
\onlinecite{motda,gren}]. This shows how a kinetic approach may be properly
generalized to AS.

\subsubsection{High temperature approximation of the time integral $I_{BA\pm
}$}

To calculate $I_{BA\pm }(\omega )$ defined by Eq.(\ref{tij}), we change the
integration variable from $s$ to $t$: $s=t-i\beta \hbar /2$. Eq.(\ref{tij})
becomes%
\begin{equation*}
I_{BA\pm }(\omega )=\exp \{-\frac{1}{2}\sum_{\alpha }\coth \frac{\beta \hbar
\omega _{\alpha }}{2}(\theta _{\alpha }^{A})^{2}\}e^{\beta \hbar (\pm \omega
+\omega _{AB})/2}
\end{equation*}%
\begin{equation}
\int_{-\infty +i\beta \hbar /2}^{i\beta \hbar /2}dte^{it(\pm \omega +\omega
_{AB)}}\exp \{\frac{1}{2}\sum_{\alpha }(\theta _{\alpha }^{A})^{2}\csc h%
\frac{\beta \hbar \omega _{\alpha }}{2}\cos t\omega _{\alpha }\}.
\label{2b0}
\end{equation}%
If we view $t$ as a complex variable, the saddle point of $\frac{1}{2}%
\sum_{\alpha }(\theta _{\alpha }^{A})^{2}\csc h\frac{\beta \hbar \omega
_{\alpha }}{2}\cos t\omega _{\alpha }$ is at ($0,0$). Because the integrand
in Eq.(\ref{2b0}) is analytic in the whole complex-$t$ plane, we can deform
the integral path from $(-\infty +i\beta \hbar /2,0+i\beta \hbar /2]$ to a
new path C$_{1}+$C$_{2}+$C$_{3}$ crossing the saddle point ($0,0$), where C$%
_{1}$: $(-\infty +i\beta \hbar /2,-\infty +i0]$, C$_{2}$: $(-\infty ,0]$, C$%
_{3}$: $(0+i0,0+i\beta \hbar /2]$. Because of the external field and
residual interactions being adiabatically introduced\cite{pss}, the
integration along C$_{1}$ is zero. When $k_{B}T\geq \hbar \overline{\omega }$
($\overline{\omega }$ is the frequency of the first peak in phonon
spectrum), $\sum_{\alpha }(\theta _{\alpha }^{A})^{2}k_{B}T/\hbar \overline{%
\omega }$ is large. The integrals along C$_{2}$ and C$_{3}$ can be
asymptotically calculated by the Laplace method\cite{bo}. The final result
for $I_{BA\pm }$ is
\begin{subequations}
\begin{equation}
I_{BA\pm }(\omega )=i\hbar /\lambda _{BA}  \label{2b01}
\end{equation}%
\end{subequations}
\begin{equation*}
+\frac{\hbar e^{\beta \hbar (\pm \omega +\omega _{AB})/2-y_{\pm
}^{BA}-\lambda _{BA}/4k_{B}T}}{(k_{B}T\lambda _{BA})^{1/2}}[\frac{\sqrt{\pi }%
}{2}-iA(y_{\pm }^{BA})],
\end{equation*}%
where
\begin{equation}
y_{\pm }^{BA}=\frac{[\hbar (\pm \omega +\omega _{AB})]^{2}}{4\lambda
_{BA}k_{B}T},\text{ \ \ }\lambda _{BA}=\frac{1}{2}\sum_{\alpha }\hbar \omega
_{\alpha }(\theta _{\alpha }^{A})^{2},  \label{ere}
\end{equation}%
and%
\begin{equation}
A(y_{\pm })=\left\{
\begin{array}{c}
\sum_{n=0}^{\infty }\frac{y_{\pm }^{n+1/2}}{n!(2n+1)},\text{ if }y_{\pm
}\leq 1 \\
\frac{e^{y_{\pm }}}{2\sqrt{y_{\pm }}}[1+\sum_{n=1}^{\infty }\frac{(2n-1)!!}{%
2^{n}y_{\pm }^{n}}],\text{ if }y_{\pm }>1%
\end{array}%
\right. .  \label{afu}
\end{equation}%
The applicable condition for a-Si is T\TEXTsymbol{>} 232K\cite%
{bah,brod74,kam}; for a-Ge is T\TEXTsymbol{>} 115K\cite{bah,brod74,pay}.

\subsubsection{Velocity matrix elements}

Under the approximations in Sec.\ref{als} and \ref{aes}, the velocity matrix
elements in Eq.(\ref{vle}) can be obtained by changing the integration
variable from $\mathbf{r}$ to $\mathbf{r}^{\prime }=\mathbf{r}-\mathbf{R}%
_{A} $, and introducing a spherical coordinate system with $\mathbf{R}_{A}$
as the origin and $\mathbf{p}$ as polar axis. One can show that $%
v_{B_{1}A}^{x}=v_{B_{1}A}^{y}=0$, i.e. for the velocity components
perpendicular to $\mathbf{p}$, the matrix elements are zero:
\begin{equation}
v_{\perp }^{B_{1}A}=0.  \label{cba}
\end{equation}%
The matrix element of $v^{z}$ (the velocity component parallel to $\mathbf{p}
$) is

\begin{equation}
v_{\parallel }^{B_{1}A}=v_{B_{1}A}^{z}=\frac{p}{m}8\pi ^{1/2}\frac{e^{-i%
\mathbf{p}\cdot \mathbf{R}_{A}/\hbar }V^{-1/2}\xi ^{3/2}}{(1+p^{2}\xi
^{2}/\hbar ^{2})^{2}}.  \label{vba}
\end{equation}%
Similarly,%
\begin{equation}
w_{\parallel }^{AB_{1}}=w_{AB_{1}}^{z}=-\frac{p}{m}8\pi ^{1/2}\frac{e^{-i%
\mathbf{p}\cdot \mathbf{R}_{A}/\hbar }V^{-1/2}\xi ^{3/2}}{(1+p^{2}\xi
^{2}/\hbar ^{2})^{2}}.  \label{wab}
\end{equation}%
Substitute Eqs.(\ref{vba},\ref{wab}) into Eq.(\ref{mef}), and the inverse of
the effective mass tensor becomes%
\begin{equation}
(m_{eff}^{B_{1}A-1})_{\alpha \beta }=\frac{m^{-1}p_{\alpha }p_{\beta }}{%
m(E_{B_{1}}^{0}-E_{A}^{0})}\frac{32\pi \xi ^{3}V^{-1}}{(1+p^{2}\xi
^{2}/\hbar ^{2})^{4}}.  \label{mom}
\end{equation}

Since for each Cartesian component\cite{motda},

\begin{equation}
\langle \chi _{B}|x_{\alpha }|\phi _{A}\rangle =\frac{i\hbar \langle \chi
_{B}|v_{\alpha }|\phi _{A}\rangle }{(E_{A}-E_{B})},\text{ \ }\alpha =x,y,z,
\label{rpm}
\end{equation}%
from (\ref{cba}) and (\ref{vba}), one has%
\begin{equation}
\langle \chi _{B}|r_{\perp }|\phi _{A}\rangle =0\text{ and }\langle \chi
_{B}|r_{\parallel }|\phi _{A}\rangle =\frac{i\hbar v_{\parallel }^{BA}}{%
(E_{A}-E_{B})}\text{.}  \label{rpc}
\end{equation}
\begin{figure}[th]
\centering
\subfigure[]{\includegraphics[scale=0.15]{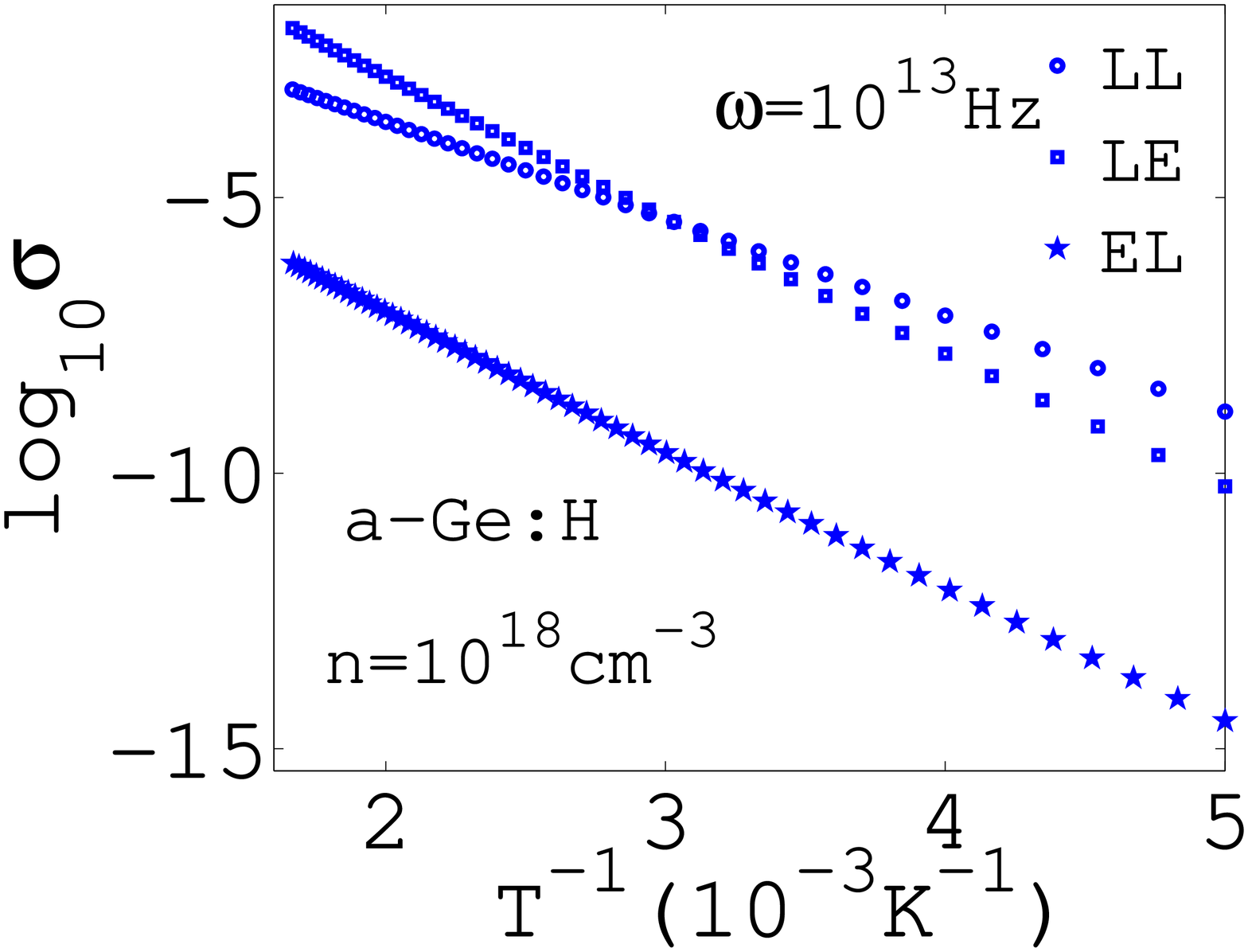}\label{Ge1813c}}\hfill %
\subfigure[]{\includegraphics[scale=0.15]{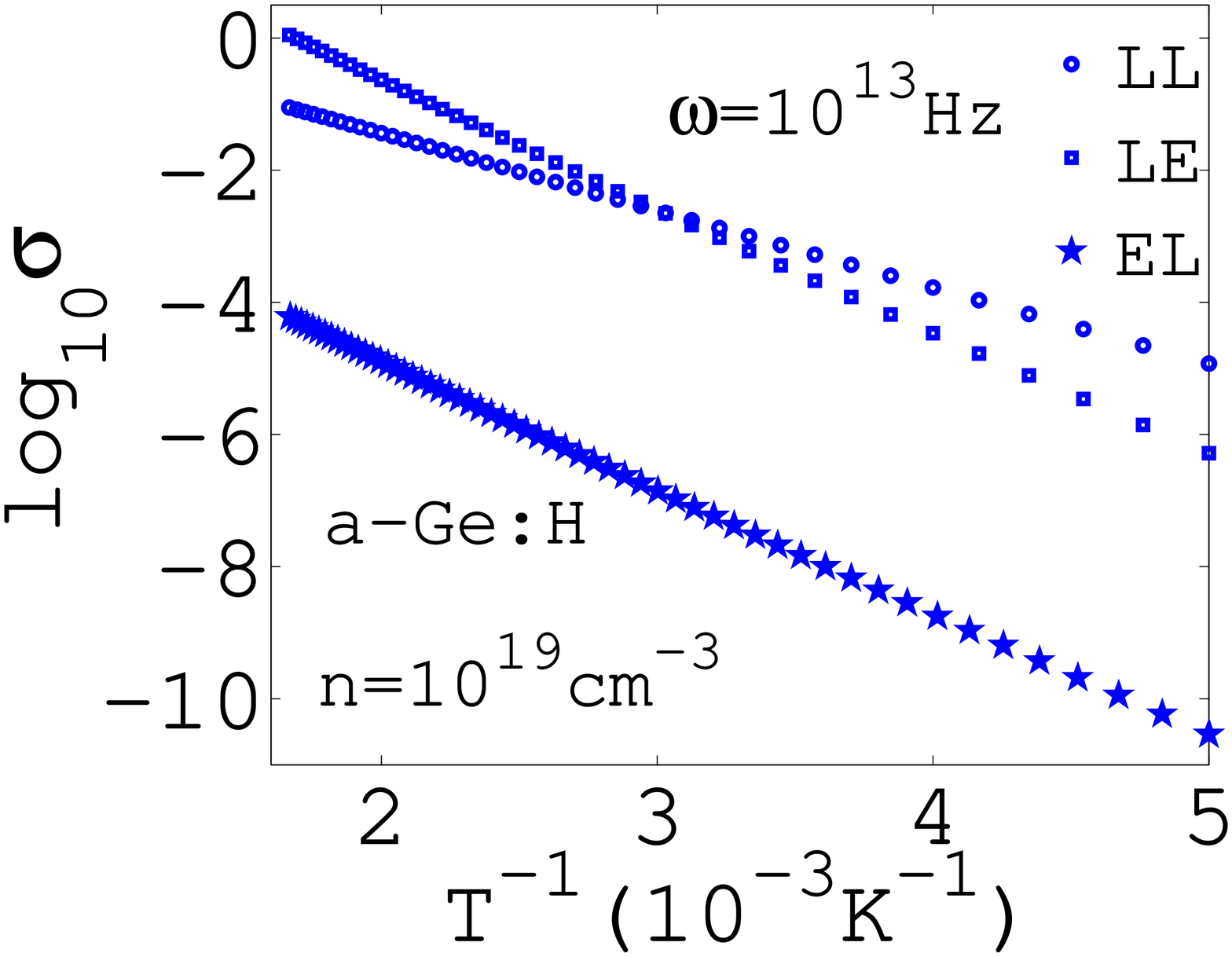}\label{Ge1913c}} \hfill %
\subfigure[]{\includegraphics[scale=0.15]{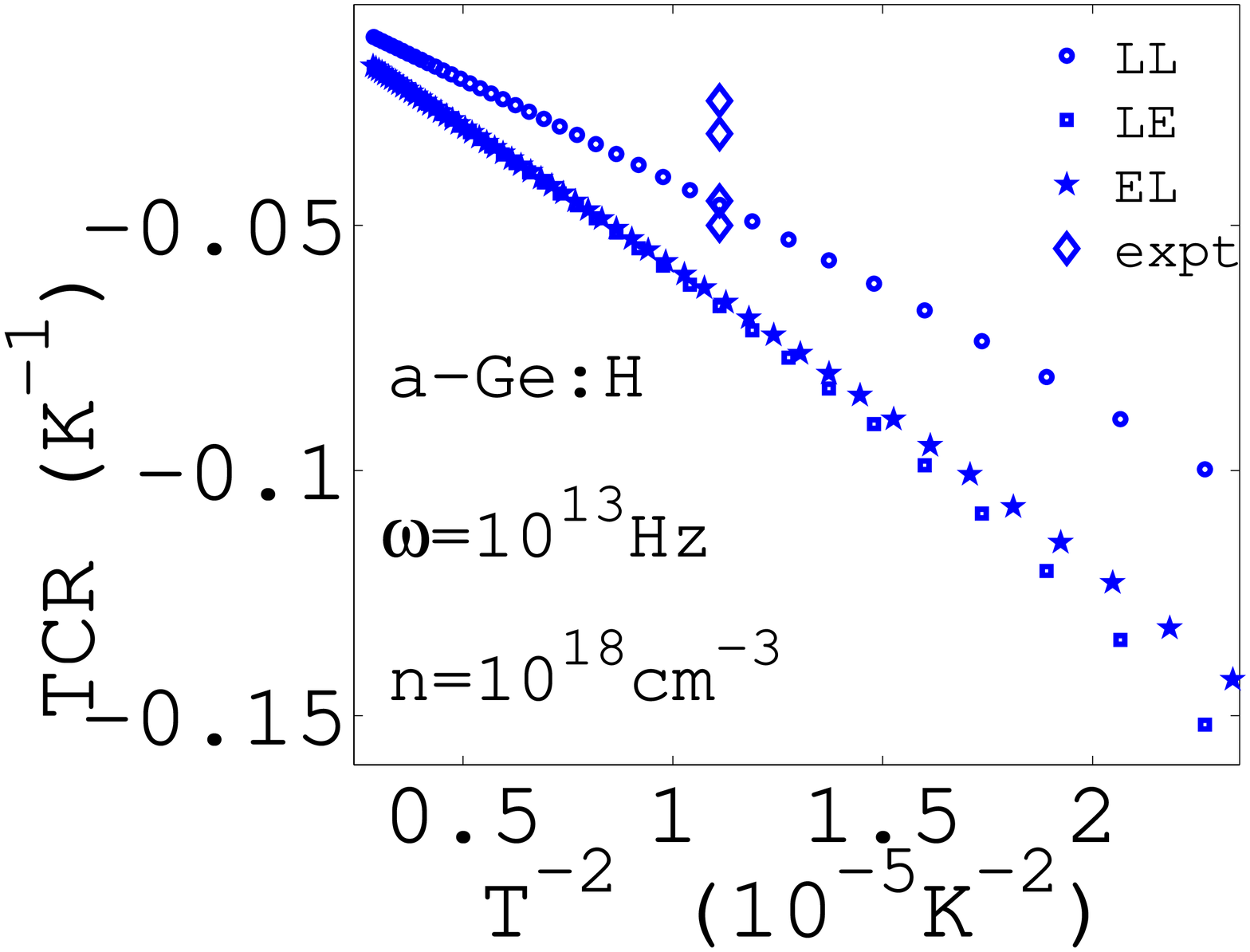}\label{Ge1813tcr}} \hfill %
\subfigure[]{\includegraphics[scale=0.15]{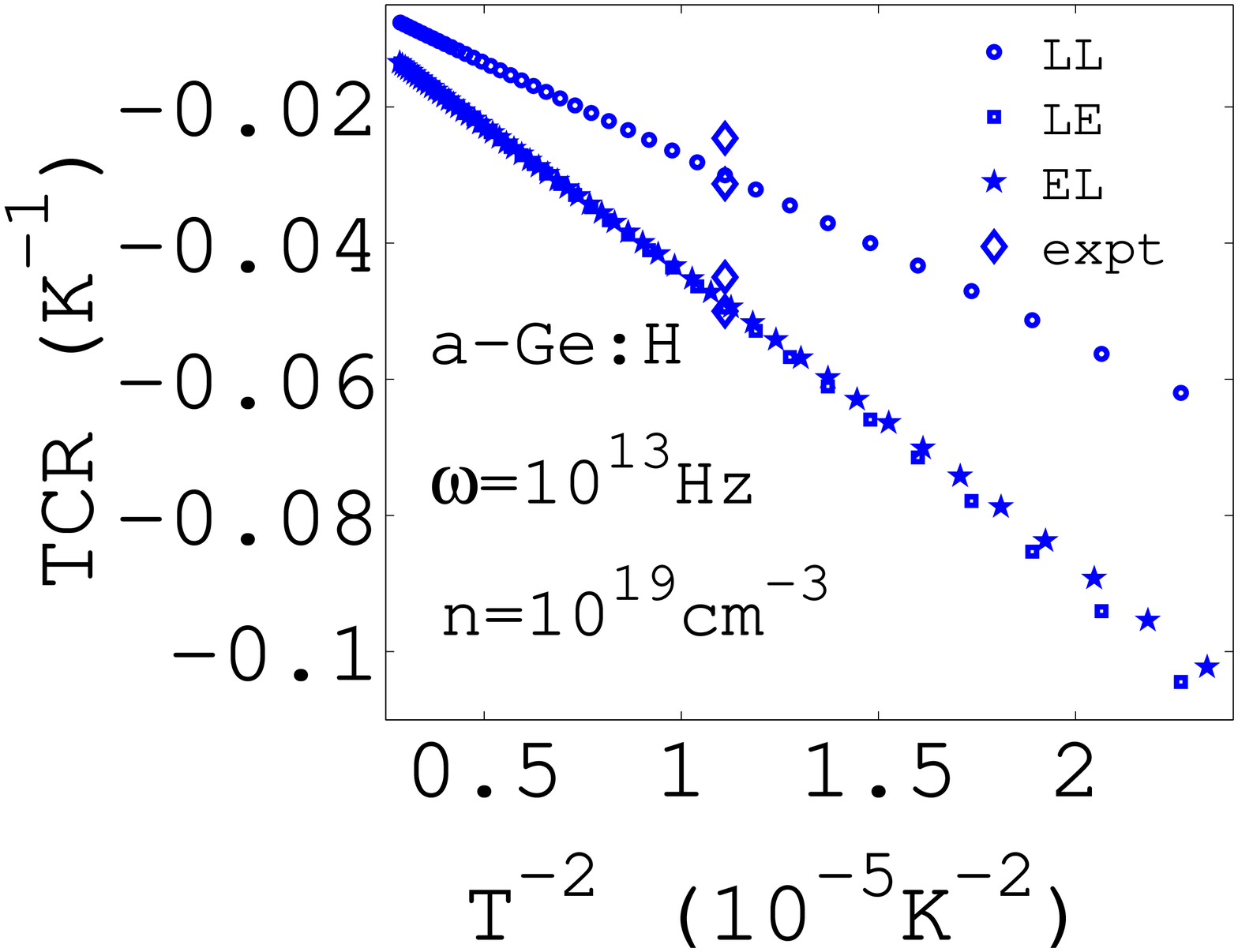}\label{Ge1913tcr}}\hfill
\caption{Conductivity and TCR as functions of temperature in two n-doped
a-Ge:H samples at $\protect\omega =10^{13}$ Hz. The experimental values are
taken from [\onlinecite{tor,ino}].}
\label{Ge13c}
\end{figure}

\subsubsection{Relation to kinetic method}

Because $\phi _{A}$ vanishes at $x_{\alpha }=\pm \infty $ ($\alpha =x,y,z$),
by means of partial integration, one can show that $w_{AB}^{\alpha
}=-v_{BA}^{\alpha }$. Then%
\begin{equation}
(w_{AB}^{\alpha }-v_{BA}^{\alpha })(v_{BA}^{\beta })^{\ast
}=-2v_{BA}^{\alpha }(v_{BA}^{\beta })^{\ast }=-\frac{2}{3}v_{BA}^{\alpha
}(v_{BA}^{\alpha })^{\ast }\delta _{\alpha \beta },  \label{dch}
\end{equation}%
the last step is correct only for a cubic or isotropic body. For such a
body, the product of two matrix elements is a real number. From the
requirement that $\func{Re}\sigma _{\alpha \beta }$ and $\func{Im}\sigma
_{\alpha \beta }$ are real numbers, we only require
\begin{equation}
\func{Re}[I_{BA+}\pm I_{BA-}]=\frac{\sqrt{\pi }\hbar }{2(k_{B}T\lambda
_{BA})^{1/2}}  \label{sbu}
\end{equation}%
\begin{equation*}
\lbrack e^{-\frac{\lambda _{BA}}{4k_{B}T}[1+\frac{(\hbar \omega _{BA}-\hbar
\omega )}{\lambda _{BA}}]^{2}}\pm e^{-\frac{\lambda _{BA}}{4k_{B}T}[1+\frac{%
(\hbar \omega _{BA}+\hbar \omega )}{\lambda _{BA}}]^{2}}]
\end{equation*}%
in expression (\ref{lec}). The temperature dependence (\ref{sbu}) is the
same as that obtained from the kinetic method\cite{epjb}, although the
Landau-Peierls condition is \textit{not} satisfied. This is a coincidence
caused by two factors. First, for LE, EL, LL and EE transitions driven by
external field, the contribution to conductivity has the form of Eq.(\ref%
{lec}). Thus only the real part of the one dimensional time integral plays a
role. In contrast to Eq.(\ref{tij}), in the corresponding kinetic expression%
\cite{epjb}, the upper limit of time integral is $\infty $ (long time limit)
rather than 0. Second, because in both cases we apply an asymptotic
expansion to calculate the time integral at high temperature, at leading
order, the real part of (\ref{tij}) is half the corresponding time integral
in kinetic theory. The difference in temperature dependence only appears in
subdominant terms.

\begin{figure}[th]
\centering
\subfigure[]{\includegraphics[scale=0.15]{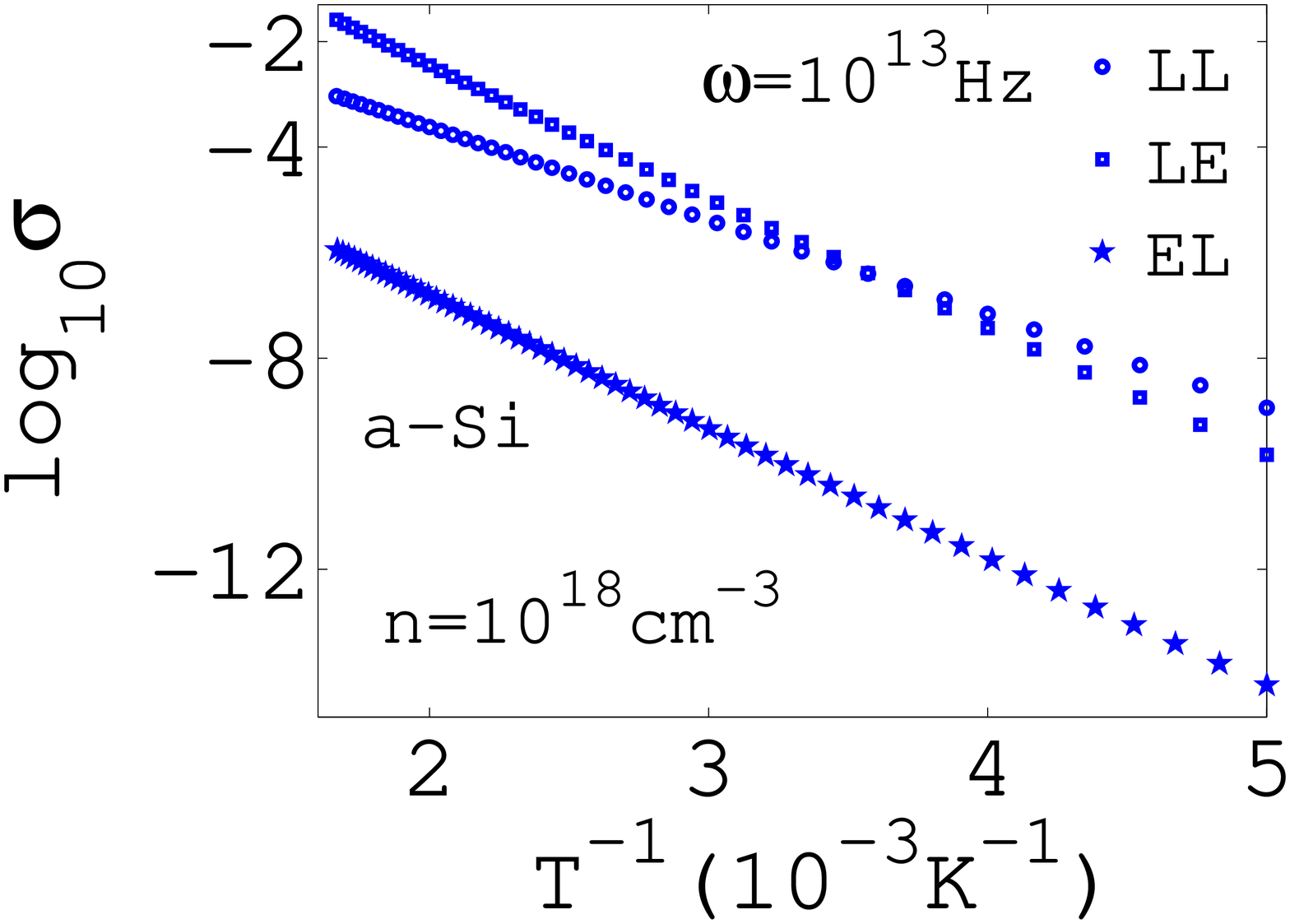}\label{1813c}}\hfill %
\subfigure[]{\includegraphics[scale=0.15]{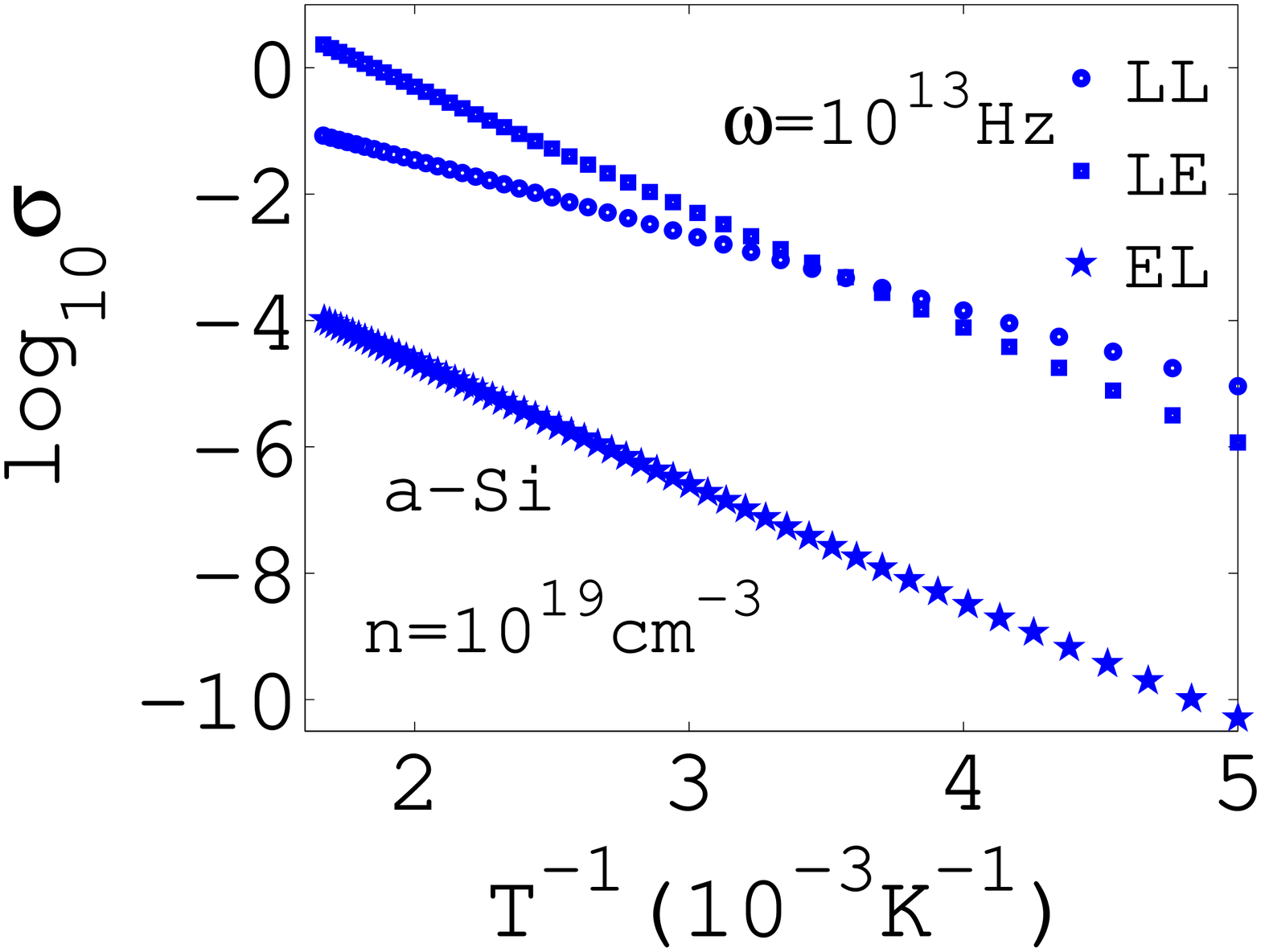}\label{1913c}}\hfill %
\subfigure[]{\includegraphics[scale=0.15]{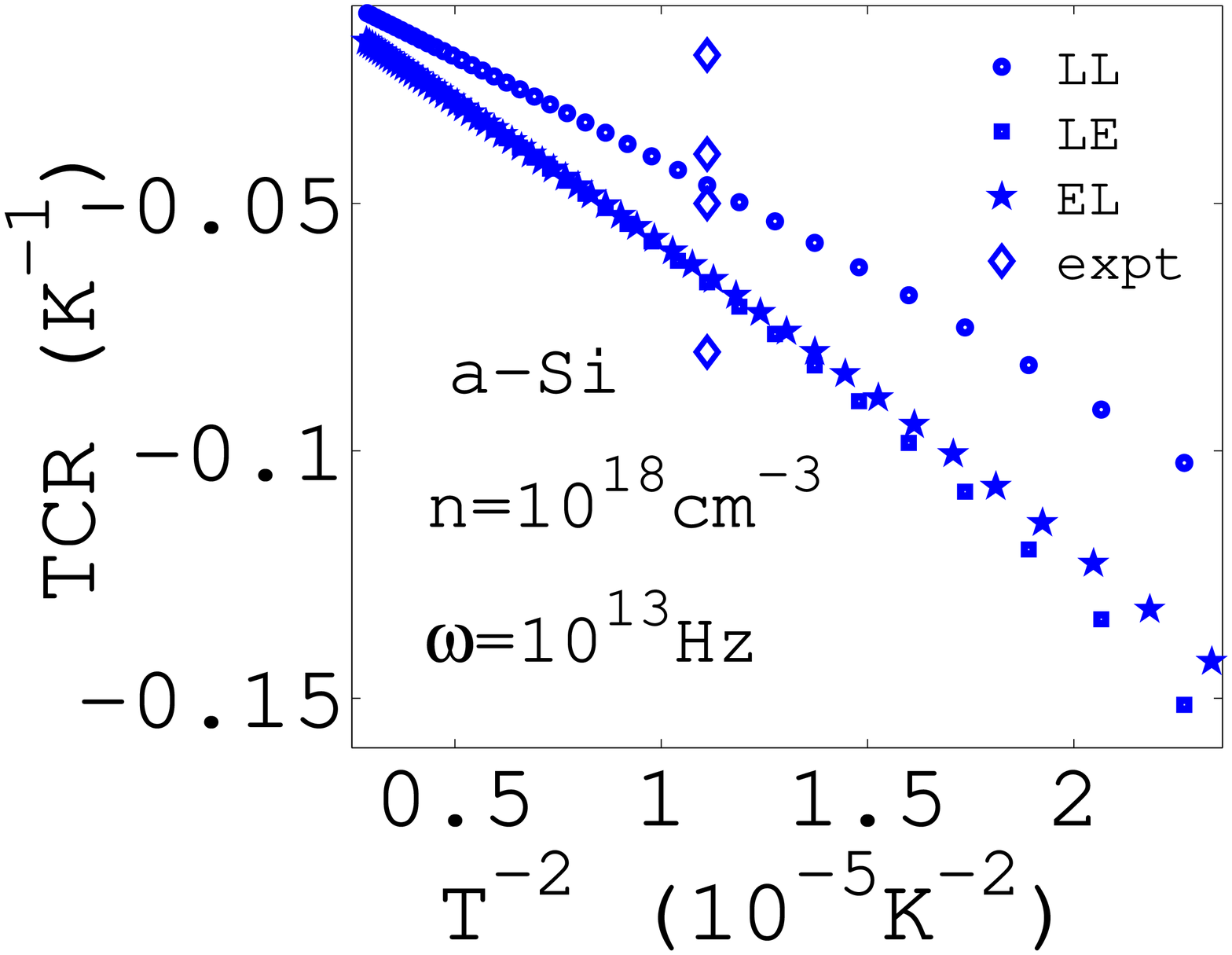}\label{1813tcr}}\hfill %
\subfigure[]{\includegraphics[scale=0.15]{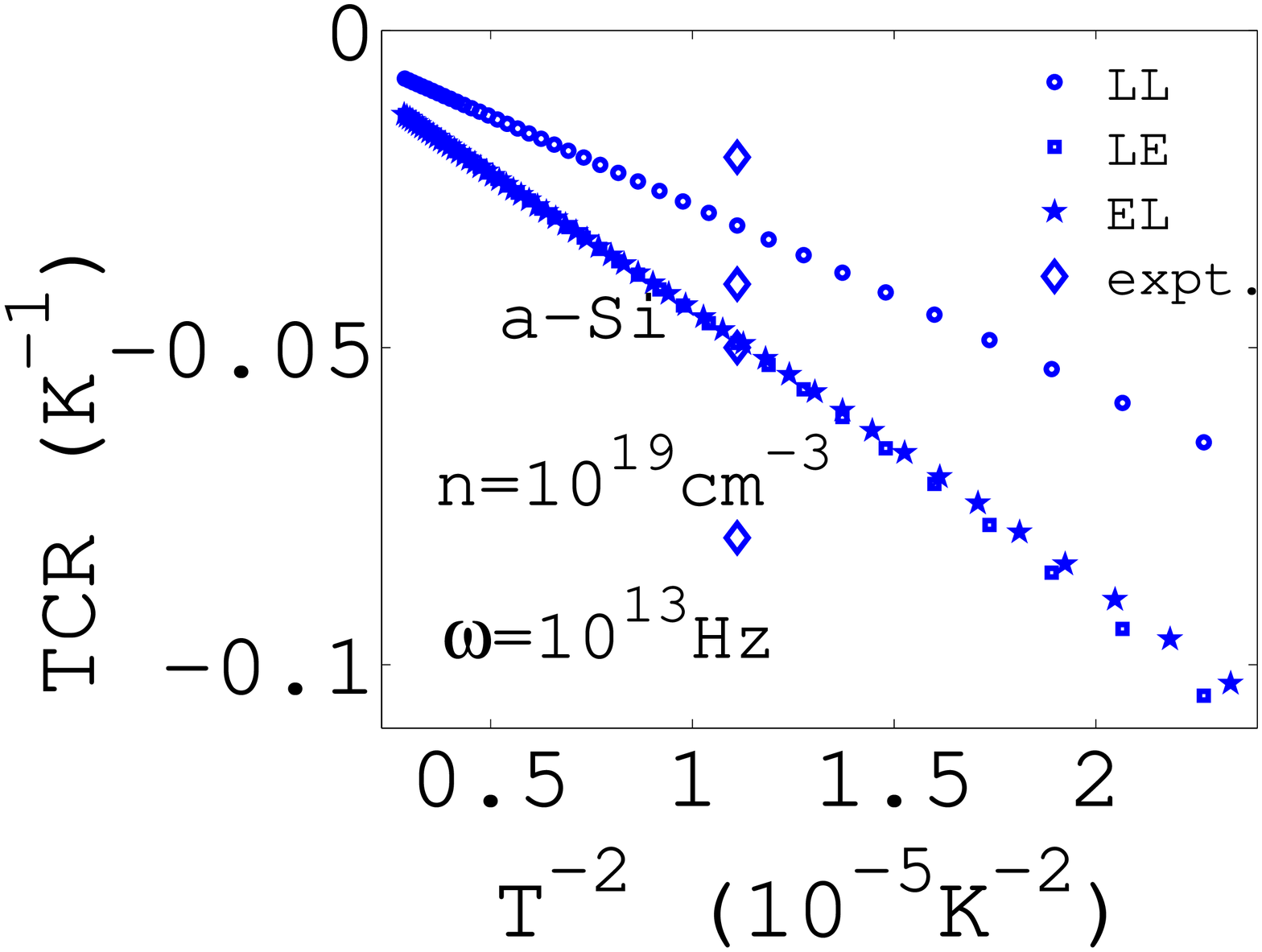}\label{1913tcr}}\hfill
\caption{Conductivity and TCR as functions of temperature in two n-doped
a-Si:H samples at $\protect\omega =10^{13}$ Hz. The experimental values are
taken from [\onlinecite{sai}]}
\label{13c}
\end{figure}

When transfer integrals or e-ph interaction are involved at first order,
various transport processes are the same order of magnitude as the processes
discussed here (zero-order in residual interaction). In these first-order
processes, it is the imaginary part of a two-fold time integral that
contribute to conductivity cf. [\onlinecite{pss}]. Some of these first order
processes do not appear in kinetic models. Even for the processes expected
from kinetic theory, the temperature dependence derived in the MRM is
different from that derived from kinetic theory.

\subsubsection{Summation over electronic states}

To carry out the sum over the final extended states and average over initial
localized states, we first carry out $\sum_{B}$ for a fixed localized state $%
A$. We take the center $\mathbf{R}_{A}$ of $\phi _{A}(\mathbf{r}-\mathbf{R}%
_{A})$ as the origin of coordinates, the incident direction $\mathbf{k/|k|}$
of electromagnetic wave as polar axis (z axis), the directions ($\mathbf{%
\epsilon }_{1},\mathbf{\epsilon }_{2}$) of two linear polarization vectors
as x and y axis respectively. The incident field is expressed as
\begin{equation}
\mathbf{F}=F_{1}\mathbf{\epsilon }_{1}+F_{2}\mathbf{\epsilon }_{2}+0\mathbf{%
k/|k|}.  \label{fdi}
\end{equation}%
Consider an extended state (a wave packet propagating along $\mathbf{p}$) $%
V^{-1/2}e^{i\mathbf{p}\cdot \mathbf{r}/\hbar }$, here for simplicity we
neglected other waves with wave vectors close to $\mathbf{p}$. We can select
an orthogonal frame ($\mathbf{l}$, $\mathbf{m}$, $\mathbf{n}$), where $%
\mathbf{n}=\mathbf{p}/|\mathbf{p}|$, $\mathbf{l}$ and $\mathbf{m}$ are two
unit vectors perpendicular to each other and perpendicular to $\mathbf{n}$.
The position vector $\mathbf{r}$ of electron can be resolved as
\begin{equation}
\mathbf{r}=r_{\perp 1}\mathbf{l}+r_{\perp 2}\mathbf{m}+r_{\parallel }\mathbf{%
n}.  \label{rso}
\end{equation}%
According to Eq.(\ref{rpc}), one has%
\begin{equation}
\langle \chi _{B}|\mathbf{r}|\phi _{A}\rangle =\mathbf{n}\langle \chi
_{B}|r_{\parallel }|\phi _{A}\rangle .  \label{pare}
\end{equation}%
The matrix elements of the perturbation of external field is simplified to%
\begin{equation}
\langle \chi _{B}|\mathbf{F}\cdot \mathbf{r}|\phi _{A}\rangle =\mathbf{F}%
\cdot \mathbf{n}\langle \chi _{B}|r_{\parallel }|\phi _{A}\rangle .
\label{fp}
\end{equation}%
\begin{equation*}
=\sin \theta (F_{1}\cos \phi +F_{2}\sin \phi )\frac{i\hbar v_{\parallel
}^{BA}}{(E_{A}-E_{B})},
\end{equation*}%
where $\theta $ is the inclination angle of $\mathbf{p}$ relative to $%
\mathbf{k}$, $\phi $ is the azimuth angle of the orthogonal projection of $%
\mathbf{p}$ on plane ($\mathbf{\epsilon }_{1},\mathbf{\epsilon }_{2}$)
relative to $\mathbf{\epsilon }_{1}$. In this coordinate system,%
\begin{equation}
\sum_{B_{1}}\rightarrow \frac{V}{(2\pi \hbar )^{3}}\int_{0}^{\infty
}dpp^{2}\int_{0}^{\pi }d\theta \sin \theta \int_{0}^{2\pi }d\phi .
\label{psu}
\end{equation}

The incident field (\ref{fdi}) has only $x$ and $y$ components. So that only
the xx, xy, yx and yy components of the conductivity tensor are involved in
the conduction process driven by field (\ref{fdi}). In consonance with Eq.(%
\ref{fp}), one should make the substitution%
\begin{equation}
v_{BA}^{x}\rightarrow v_{\parallel }^{BA}\sin \theta \cos \phi ,\text{ \ }%
v_{BA}^{y}\rightarrow v_{\parallel }^{BA}\sin \theta \sin \phi ,  \label{bus}
\end{equation}%
in the conductivity tensor (\ref{lec}). The angular part of integral (\ref%
{psu}) can be carried out. From Eqs.(\ref{psu},\ref{bus}), one can see $%
\sigma _{xy}=\sigma _{yx}=0$ and $\sigma _{xx}=\sigma _{yy}=\sigma $.
Because the factors in Eq.(\ref{lec}) do not depend on the position of
localized state $\phi _{A}$, one can carry out the spatial integral in $%
\sum_{A}$. The conductivity from LE transitions is%
\begin{equation*}
\left\{
\begin{array}{c}
\func{Re} \\
\func{Im}%
\end{array}%
\right. \sigma (\omega )=\frac{4\pi \overline{\xi }^{3}}{3}\frac{%
bZe^{2}n_{loc}}{4\pi \epsilon _{0}\varepsilon U}\frac{8ne^{2}}{3\pi \hbar
^{3}m^{2}}\int_{0}^{\infty }d\xi \int_{0}^{\infty }dp
\end{equation*}%
\begin{equation}
\lbrack 1-f(E_{B_{1}})]f(E_{A})\frac{p^{4}}{(E_{B_{1}}^{0}-E_{A}^{0})}\frac{%
\xi \exp (-\frac{bZe^{2}}{4\pi \epsilon _{0}\varepsilon U\xi })}{(1+p^{2}\xi
^{2}/\hbar ^{2})^{4}}  \label{cle}
\end{equation}%
\begin{equation*}
\frac{\sqrt{\pi }\hbar }{2(k_{B}T\lambda _{BA})^{1/2}}[e^{-\frac{\lambda
_{BA}}{4k_{B}T}(1+\frac{\hbar \omega _{BA}-\hbar \omega }{\lambda _{BA}}%
)^{2}}\pm e^{-\frac{\lambda _{BA}}{4k_{B}T}(1+\frac{\hbar \omega _{BA}+\hbar
\omega }{\lambda _{BA}})^{2}}],
\end{equation*}%
where $n=N_{e}/\Omega _{\mathbf{s}}$ is the carrier concentration, $E_{A}$
and $E_{B_{1}}$ are given in Eqs.(\ref{lh},\ref{ted}). From Eq.(\ref{cle}),
one can easily compute TCR: $\rho ^{-1}\frac{d\rho }{dT}=-\sigma ^{-1}\frac{%
d\sigma }{dT}$, an important material parameter for bolometer\cite{str,sai}.
$\sigma $ and TCR are expressed with easy access quantities: $U$ and $E_{c}$
for localized states, $\varepsilon $ and $q_{TF}$ for the interaction
between electron and atomic core, the averaged sound speed $\overline{c}$
for the vibrations.

\subsection{EL transitions driven by external field}

\label{el}

Since the field-matter coupling is Hermitian, the corresponding expressions
for EL transition driven by field can be obtained from those for LE
transitions driven by field through exchanging the status of $\phi _{A}$ and
$\chi _{B}$.%
\begin{equation*}
\left\{
\begin{array}{c}
\func{Re} \\
\func{Im}%
\end{array}%
\right. \sigma (\omega )=\frac{4\pi \overline{\xi }^{3}}{3}\frac{%
bZe^{2}n_{loc}}{4\pi \epsilon _{0}\varepsilon U}\frac{8ne^{2}}{3\pi \hbar
^{3}m^{2}}\int_{0}^{\infty }d\xi \int_{0}^{\infty }dp
\end{equation*}%
\begin{equation}
\lbrack 1-f(E_{A})]f(E_{B})\frac{p^{4}}{(E_{A}^{0}-E_{B}^{0})}\frac{\xi \exp
(-\frac{bZe^{2}}{4\pi \epsilon _{0}\varepsilon U\xi })}{(1+p^{2}\xi
^{2}/\hbar ^{2})^{4}}  \label{cel1}
\end{equation}%
\begin{equation*}
\frac{\sqrt{\pi }\hbar }{2(k_{B}T\lambda _{AB})^{1/2}}[e^{-\frac{\lambda
_{AB}}{4k_{B}T}(1+\frac{\hbar \omega _{AB}-\hbar \omega }{\lambda _{AB}}%
)^{2}}\pm e^{-\frac{\lambda _{AB}}{4k_{B}T}(1+\frac{\hbar \omega _{AB}+\hbar
\omega }{\lambda _{AB}})^{2}}],
\end{equation*}%
where%
\begin{equation}
y_{\pm }^{AB}=\frac{(\omega _{BA}\pm \omega )^{2}}{4\lambda _{AB}k_{B}T},%
\text{ \ \ }\lambda _{AB}=\frac{1}{2}\sum_{\alpha }\hbar \omega _{\alpha
}(\theta _{\alpha }^{A})^{2}.  \label{yab}
\end{equation}%
For the LE transition driven by the transfer integral and the EL transition
driven by e-ph interaction, one does not have this symmetry\cite{epjb,pss}.


\subsection{LL transition driven by external field}

\label{ll}

One can similarly find the conductivity from the LL transitions driven by
external field (Fig.2a of [\onlinecite{pss}]):

\begin{equation*}
\left\{
\begin{array}{c}
\func{Re} \\
\func{Im}%
\end{array}%
\right. \sigma _{\alpha \beta }(\omega )=-\frac{N_{e}e^{2}}{2\Omega _{s}}%
\sum_{AA_{1}}\func{Im}\frac{(w_{AA_{1}}^{\alpha }-v_{A_{1}A}^{\alpha
})(v_{A_{1}A}^{\beta })^{\ast }}{(E_{A}^{0}-E_{A_{1}}^{0})}
\end{equation*}%
\begin{equation}
i[I_{A_{1}A+}\pm I_{A_{1}A-}][1-f(E_{A_{1}})]f(E_{A}),  \label{cll}
\end{equation}%
where the velocity matrix elements are%
\begin{equation}
w_{AA_{1}}^{\alpha }=-\frac{i\hbar }{m}\int d^{3}x\phi (\mathbf{r}-\mathbf{R}%
_{A})\frac{\partial }{\partial x_{\alpha }}\phi ^{\ast }(\mathbf{r}-\mathbf{R%
}_{A_{1}}),  \label{waa1}
\end{equation}%
and%
\begin{equation}
v_{A_{1}A}^{\alpha }=-\frac{i\hbar }{m}\int d^{3}x\phi ^{\ast }(\mathbf{r}-%
\mathbf{R}_{A_{1}})\frac{\partial }{\partial x_{\alpha }}\phi (\mathbf{r}-%
\mathbf{R}_{A}).  \label{va1a}
\end{equation}%
$v_{A_{1}A}^{\alpha }$ is given in Eq.(\ref{vz}) and $w_{AA_{1}}^{\alpha
}=-v_{A_{1}A}^{\alpha }$. The time integral%
\begin{equation}
I_{A_{1}A\pm }(\omega )=\exp \{-\frac{1}{2}\sum_{\alpha }(\theta _{\alpha
}^{A_{1}}-\theta _{\alpha }^{A})^{2}\coth \frac{\beta \hbar \omega _{\alpha }%
}{2}\}  \label{r0+}
\end{equation}%
\begin{equation*}
\int_{-\infty }^{0}dse^{\pm i\omega s}e^{-is(E_{A_{1}}^{\prime
}-E_{A}^{\prime })/\hbar }
\end{equation*}%
\begin{equation*}
\exp [\frac{1}{2}\sum_{\alpha }\frac{(\theta _{\alpha }^{A_{1}}-\theta
_{\alpha }^{A})^{2}}{2}(\coth \frac{\beta \hbar \omega _{\alpha }}{2}\cos
\omega _{\alpha }s+i\sin \omega _{\alpha }s)],
\end{equation*}%
contains the primary temperature dependence of conductivity. At high
temperature $k_{B}T\geq \hbar \overline{\omega }$, $I_{A_{1}A\pm }$ reduces
to

\begin{equation}
I_{A_{1}A\pm }(\omega )=-i\hbar /\lambda _{A_{1}A}  \label{2a00}
\end{equation}%
\begin{equation*}
+\frac{\hbar e^{-\beta \hbar (\pm \omega +\omega _{AA_{1}})/2-y_{\pm
}^{A_{1}A}-\beta \lambda _{A_{1}A}/4}}{(\lambda _{A_{1}A}k_{B}T)^{1/2}}[%
\frac{\sqrt{\pi }}{2}-iA(y_{\pm }^{A_{1}A})],
\end{equation*}%
where%
\begin{equation}
\lambda _{A_{1}A}=\frac{1}{2}\sum_{\alpha }\hbar \omega _{\alpha }(\theta
_{\alpha }^{A_{1}}-\theta _{\alpha }^{A})^{2},  \label{rea1}
\end{equation}%
and%
\begin{equation}
y_{\pm }^{A_{1}A}=\frac{[\hbar (\pm \omega +\omega _{AA_{1}})]^{2}}{4\lambda
_{A_{1}A}k_{B}T}.  \label{yL}
\end{equation}

To carry out the summation over initial and final electronic states, we
first fix the initial electronic state $A$. We take the center $\mathbf{R}%
_{A}$ of localized state $\phi _{A}(\mathbf{r}-\mathbf{R}_{A})$ as the
origin, the incident direction $\mathbf{k}$ of the electromagnetic wave as
the polar axis. Denote $R=R_{AA_{1}}=|\mathbf{R}_{A_{1}}-\mathbf{R}_{A}|$
the distance between the centers of localized states $\phi _{A_{1}}(\mathbf{r%
}-\mathbf{R}_{A_{1}})$ and $\phi _{A}$, the unit vector along $(\mathbf{R}%
_{A_{1}}-\mathbf{R}_{A})$ is $\mathbf{n}%
_{AA_{1}}=(X_{AA_{1}},Y_{AA_{1}},Z_{AA_{1}})/R_{AA_{1}}$, where $%
(X_{AA_{1}},Y_{AA_{1}},Z_{AA_{1}})$ are the Cartesian components of vector $%
\mathbf{R}_{A_{1}}-\mathbf{R}_{A}$.
\begin{figure}[th]
\centering
\subfigure[]{\includegraphics[scale=0.15]{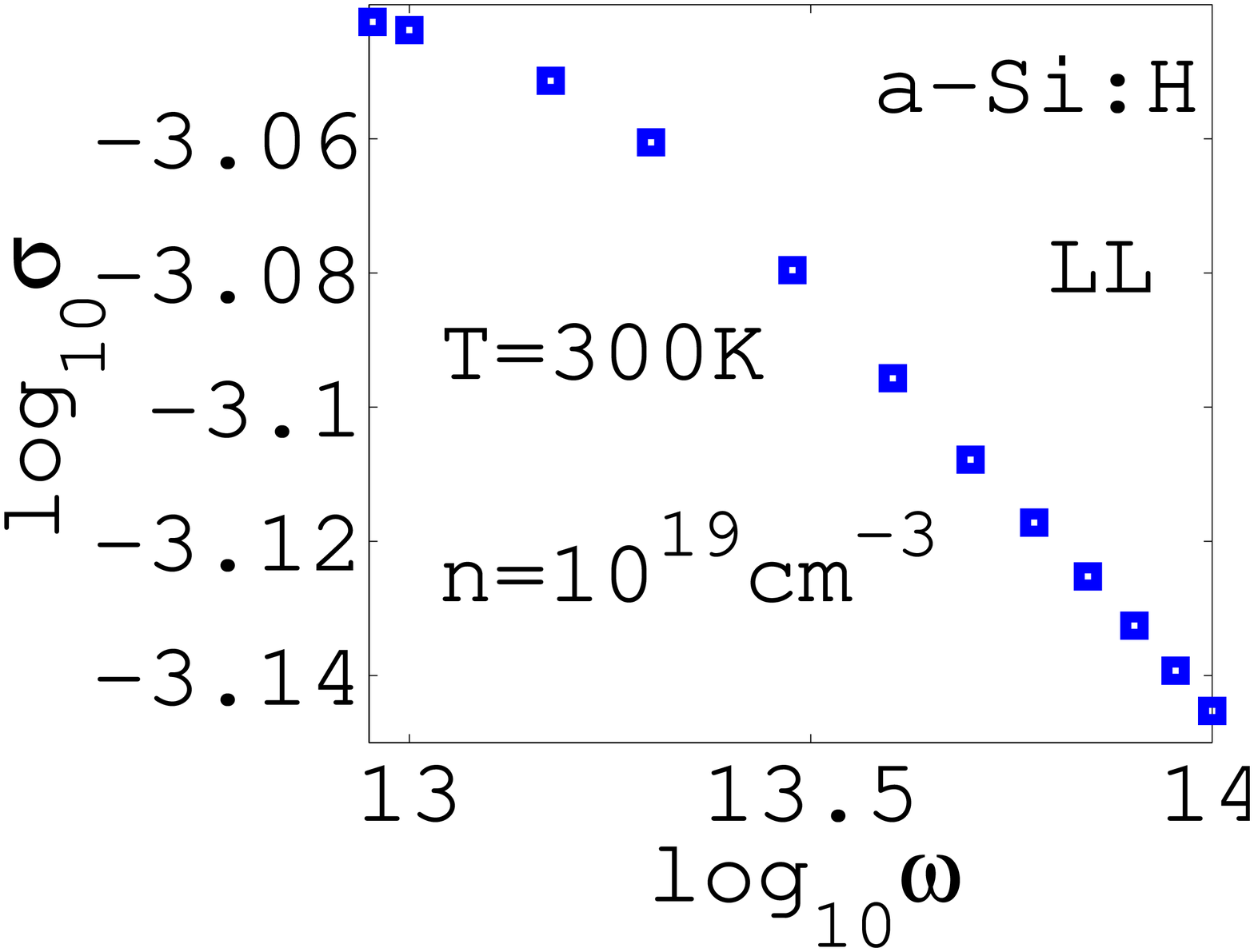}\label{fLL}}\hfill %
\subfigure[]{\includegraphics[scale=0.15]{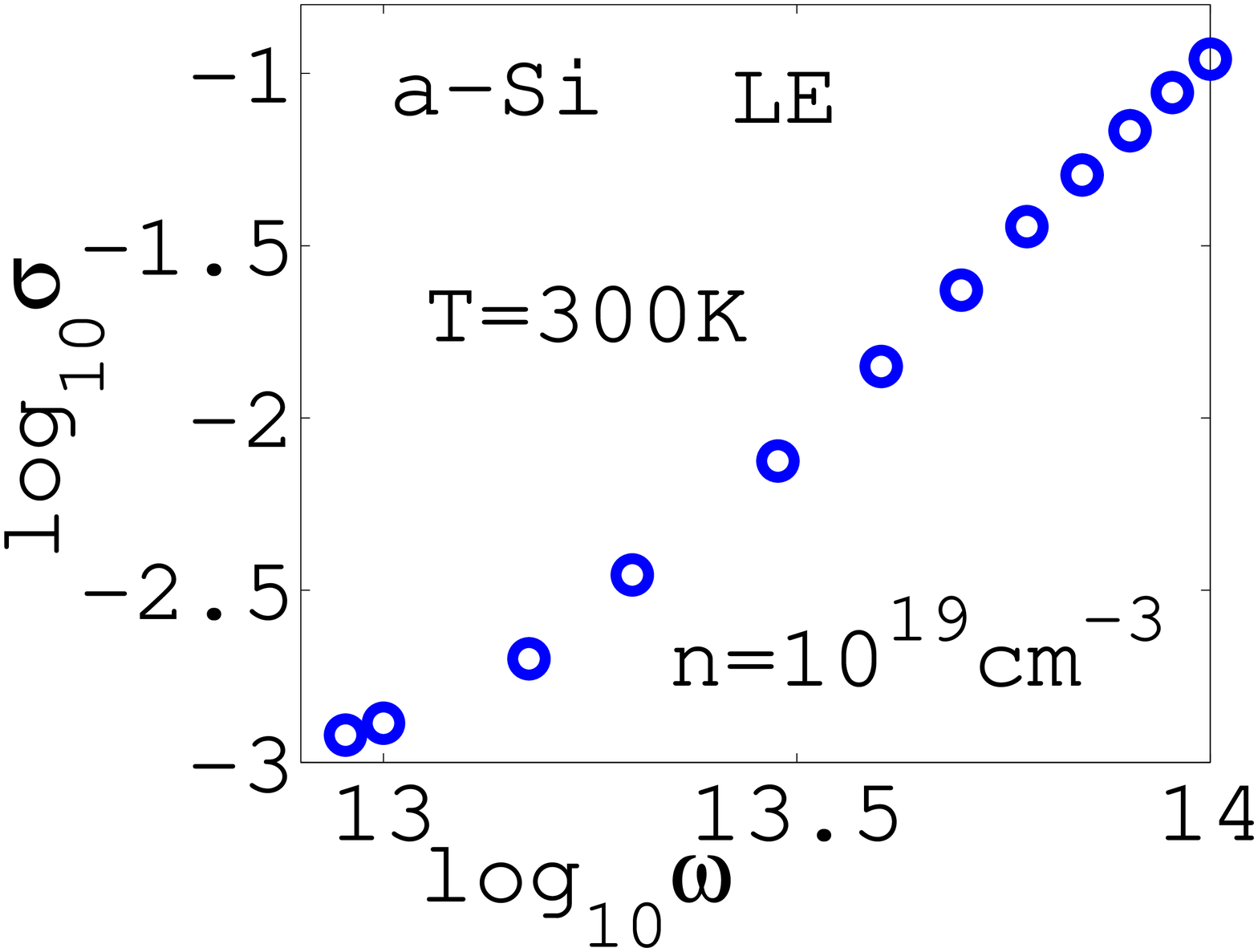}\label{fLE}}\hfill
\caption{Conduction as function of frequency at T=300K. \protect\ref{fLL}:
LL transition; \protect\ref{fLE}:LE transition }
\label{fcon}
\end{figure}

Since the conductivity tensor is usually expressed in a system of Cartesian
coordinates, we introduce an auxiliary Cartesian system ($\mathbf{\epsilon }%
_{1},\mathbf{\epsilon }_{2},\mathbf{k}$), where $\mathbf{\epsilon }_{1}$ and
$\mathbf{\epsilon }_{2}$ are the two linear polarization vectors. The
electric field $\mathbf{F}$ has only x and y components: $\mathbf{F}=F_{1}%
\mathbf{\epsilon }_{1}+F_{2}\mathbf{\epsilon }_{2}+0\mathbf{k}$. Because we
sum over $A_{1}$, the centers $\mathbf{R}_{A_{1}}$ of localized states $\phi
_{A_{1}}(\mathbf{r}-\mathbf{R}_{A_{1}})$ sit at different points. To
simplify the calculation of the velocity matrix elements, we resolve the
position vector $\mathbf{r}$ of electron in an orthogonal frame:

\begin{equation}
\mathbf{r}=r_{\perp 1}\mathbf{l}_{AA_{1}}+r_{\perp 2}\mathbf{m}%
_{AA_{1}}+r_{\parallel }\mathbf{n}_{AA_{1}},  \label{oth}
\end{equation}%
where $\mathbf{l}_{AA_{1}}$ and $\mathbf{m}_{AA_{1}}$ are two unit vectors
perpendicular to each other and to $\mathbf{n}_{AA_{1}}$. From Eq.(\ref{1rv}%
), one has%
\begin{equation}
\langle \phi _{A_{1}}|\mathbf{r}|\phi _{A}\rangle =\mathbf{n}%
_{AA_{1}}\langle \phi _{A_{1}}|r_{\parallel }|\phi _{A}\rangle .  \label{pel}
\end{equation}%
By means of Eq.(\ref{pel}), the perturbation of the electric field is%
\begin{equation}
\langle \phi _{A_{1}}|\mathbf{F}\cdot \mathbf{r}|\phi _{A}\rangle =\mathbf{F}%
\cdot \mathbf{n}_{AA_{1}}\langle \phi _{A_{1}}|r_{\parallel }|\phi
_{A}\rangle  \label{fdr}
\end{equation}%
\begin{equation*}
=\sin \theta (F_{1}\cos \phi +F_{2}\sin \phi )\frac{i\hbar v_{\parallel
}^{A_{1}A}}{(E_{A}-E_{A_{1}})},
\end{equation*}%
where $v_{\parallel }^{A_{1}A}$ has been obtained in appendix \ref{vm}. The
angular integrals in summation $\sum_{A_{1}}$ can be effected: $\sigma
_{xy}=\sigma _{yx}=0$ and $\sigma _{xx}=\sigma _{yy}=\sigma $. Because of
the uniformity of AS, the spatial integral in $\sum_{A}$ can be carried out.
The conductivity from LL transition driven by field is%
\begin{equation*}
\left\{
\begin{array}{c}
\func{Re} \\
\func{Im}%
\end{array}%
\right. \sigma (\omega )=\frac{4\pi \overline{\xi }^{3}}{3}[\frac{%
bZe^{2}n_{loc}}{4\pi \epsilon _{0}\varepsilon U}]^{2}\int_{0}^{\infty }\frac{%
d\xi _{1}}{\xi _{1}^{2}}\exp (-\frac{bZe^{2}}{4\pi \epsilon _{0}\varepsilon
U\xi _{1}})
\end{equation*}%
\begin{equation*}
\int_{0}^{\infty }\frac{d\xi _{2}}{\xi _{2}^{2}}\exp (-\frac{bZe^{2}}{4\pi
\epsilon _{0}\varepsilon U\xi _{2}})ne^{2}\frac{[1-f(E_{A_{1}})]f(E_{A})}{%
2(E_{A}^{0}-E_{A_{1}}^{0})}
\end{equation*}%
\begin{equation*}
\frac{\sqrt{\pi }\hbar }{2(\lambda _{A_{1}A}k_{B}T)^{1/2}}[e^{-\frac{\lambda
_{A_{1}A}}{4k_{B}T}(1+\frac{\hbar \omega _{A_{1}A}-\hbar \omega }{\lambda
_{A_{1}A}})^{2}}\pm e^{-\frac{\lambda _{A_{1}A}}{4k_{B}T}(1+\frac{\hbar
\omega _{A_{1}A}+\hbar \omega }{\lambda _{A_{1}A}})^{2}}]
\end{equation*}%
\begin{equation}
\int_{0}^{R_{c}}R^{2}dR\frac{4\pi }{3}(w_{\parallel }^{AA_{1}}-v_{\parallel
}^{A_{1}A})(v_{\parallel }^{A_{1}A})^{\ast },  \label{ddc}
\end{equation}%
where and in appendix, to shorten the symbols, we use $\xi _{2}$ instead of $%
\xi _{A_{1}}$, use $\xi _{1}$ instead of $\xi _{A}$.

We can see from Eqs.(\ref{cle},\ref{cel1},\ref{ddc}) that when $\omega =0$, $%
\func{Im}\sigma =0$ for LL, LE and EL transitions. For two n-doped a-Ge:H
samples with $n=10^{18}$ and $10^{19}$cm$^{-3}$, $\log _{10}\sigma $ and TCR
from LL, LE and EL transitions as functions of temperature at $\omega =0$
are plotted in Fig.\ref{con}. The corresponding results at $\omega =10^{13}$
Hz are plotted in Fig. \ref{Ge13c}. $\func{Re}\sigma $ increases with
frequency while TCR decreases with frequency. For two n-doped a-Si:H
samples, the conductivity and TCR as functions of temperature at $\omega
=10^{13}$Hz are plotted in Fig.\ref{13c}, the results at $\omega =0$ was
reported in [\onlinecite{s47}]. The calculated TCR for a-Si:H falls\cite{s47}
in the observed\cite{sai,ord,kra} range between -2\% and -8\%.

At $\omega =0$, the conductivity from LE transition is the same order of
magnitude as that from LL transitions, the conductivity from EL transitions
is much smaller than those from LL and LE transitions. There is a crossover
temperature T$^{\ast }$, below T$^{\ast }$ the conductivity from LL
transitions is larger than the conductivity from LE transitions, above T$%
^{\ast }$ the conductivity from LE transitions is larger. Because the
activation energy for LL transitions is different to that for LE
transitions, this phenomenon explained the kink on the observed $\log
_{10}\sigma $ vs. 1/T curve\cite{s47}.

For two n-doped a-Si:H samples at 300K, $\log _{10}\sigma (\omega )$ vs. $%
\log _{10}\omega $ in a frequency range $10^{13}$ to $10^{14}$Hz is
illustrated in Fig. \ref{fLE}. We can see that (i) the conductivity of LL
transitions slowly decreases with $\omega $; (ii) the conductivity from LE
transitions increases rapidly with frequency. The total conductivity is a
sum from various processes\cite{pss}, and the conductivity from LL
transitions is smaller than that from the LE transitions at higher
frequency. The total conductivity arises mainly from LE transitions at
higher frequencies. The general trend in $\log _{10}\sigma (\omega )$ vs. $%
\log _{10}\omega $ is not far from Tanaka and Fan's\cite{fan} result $\sigma
(\omega )\thicksim \omega ^{2}$, but obviously deviates from the simple
power law around $10^{13}$Hz. We must be cautious that the results derived
in this work is only suitable to the contributions from electrons: at such
high frequency the ionic contribution should also be included.

\section{Conclusion}

The microscopic response method expresses transport coefficients with
transition amplitude rather than transition probability per unit time, and
may be used in amorphous semiconductors in which Landau-Peierls condition is
violated\cite{pei,pei5}.

We presented an approximate theory for the conductivity and Hall mobility in
amorphous semiconductors systematically derived from the MRM. We obtained
the temperature dependence of the conductivity from the three simplest
transitions: LL, LE and EL transitions driven solely by field, cf. Eqs.(\ref%
{cll},\ref{cle},\ref{cel1}). The conductivity is expressed in terms of
accessible physical quantities: mobility edge, Urbach energy, static
dielectric constant and elastic modulus. LE transition (ignored in previous
theories) contributes to conductivity in the same order as LL and EE
transitions. Below a crossover temperature T$^{\ast }$, the conductivity
from LL transitions is larger than that from LE transitions; above T, the
conductivity form LE transitions is larger. This phenomenon, and different
activation energy for LL and LE transitions is the reason for the kink in
the observed conductivity vs. 1/T curve. We show how a kinetic theory of
transport can be properly generalized for AS.


\begin{acknowledgements}
We thank for support from the U.S. Army Research
Laboratory and the U. S. Army Research Office under grant number  W911NF-11-1-0358 and NSF under DMR 09-03225 .
\end{acknowledgements}

\appendix

\section{velocity matrix elements between two localized states}

\label{vm}

To calculate the velocity matrix elements in Eq.(\ref{va1a}), it is
convenient to adopt a system of spherical coordinates. We take the center $%
\mathbf{R}_{A}$ of localized state $\phi _{A}(\mathbf{r}-\mathbf{R}_{A})$ as
the origin $\mathbf{R}_{A}=0$, the connection line $\mathbf{R}_{A_{1}}-%
\mathbf{R}_{A}$ between the centers of two localized states as the polar
axis. Denote $r=|\mathbf{r}-\mathbf{R}_{A}|$ and $r_{2}=|\mathbf{r}-\mathbf{R%
}_{A_{1}}|=[r^{2}+R^{2}-2rR\cos \theta ]^{1/2}$, where $R=\mathbf{R}_{A_{1}}-%
\mathbf{R}_{A}$, $\theta $ is the angle between $\mathbf{r}-\mathbf{R}_{A}$
and $\mathbf{R}_{A_{1}}-\mathbf{R}_{A}$. The $v_{z}$ matrix element can be
written as

\begin{equation*}
v_{A_{1}A}^{z}=-\frac{i\hslash }{m}\pi ^{-1}\xi _{1}^{-3/2}\xi
_{2}^{-3/2}\int_{0}^{\infty }r^{2}dr
\end{equation*}%
\begin{equation*}
\int_{0}^{\pi }\sin \theta d\theta \int_{0}^{2\pi }d\phi e^{-r_{2}/\xi _{2}}%
\frac{\partial }{\partial z}e^{-r/\xi _{1}},
\end{equation*}%
and one has similar expressions for the matrix elements of $v_{x}$ and $%
v_{y} $. Because $r_{2}$ does not depend on the azimuth angle $\phi $,
\begin{equation}
v_{A_{1}A}^{x}=v_{A_{1}A}^{y}=0.  \label{env}
\end{equation}%
We condense them as $v_{\perp }^{A_{1}A}=0$: the matrix element for any
component of velocity perpendicular to the connection line between two
localized states is zero.

The $\phi $ integral is immediate, the remaining $r$ and $\theta $ integrals
in $v_{A_{1}A}^{z}$ can be calculated by changing the integration variable $%
\theta $ to $r_{2}$ for a fixed $r$. With the help of $\cos \theta =\frac{%
r^{2}+R^{2}-r_{2}^{2}}{2rR}$ and $\sin \theta d\theta =\frac{r_{2}dr_{2}}{Rr}
$, the integral over $\theta $ becomes an integral over $r_{2}$. One first
carries out the integral over $r_{2}$, then carries out the integral over $r$%
. For the velocity component parallel to the connection line between two
localized states, the matrix element is
\begin{equation*}
v_{\parallel }^{A_{1}A}=-\frac{i\hslash }{m}\pi ^{-1}\xi _{1}^{-3/2}\xi
_{2}^{-3/2}\int d^{3}xe^{-r_{2}/\xi _{2}}\nabla _{\parallel }e^{-r/\xi _{1}}
\end{equation*}%
\begin{equation*}
=-\frac{i\hslash }{m}(\xi _{1}\xi _{2})^{-3/2}\{-4(\frac{\xi _{2}^{2}}{R^{2}}%
+\frac{\xi _{2}}{R})\frac{e^{-R/\xi _{2}}\xi ^{\prime 3}}{\xi _{1}}
\end{equation*}%
\begin{equation*}
-(2+6\frac{\xi _{2}}{R}+6\frac{\xi _{2}^{2}}{R^{2}})\frac{\xi _{2}e^{-R/\xi
_{2}}\xi ^{\prime 2}}{\xi _{1}}-(2+6\frac{\xi _{2}}{R}+6\frac{\xi _{2}^{2}}{%
R^{2}})\frac{\xi _{2}^{2}e^{-R/\xi _{2}}\xi ^{\prime }}{\xi _{1}}
\end{equation*}%
\begin{equation*}
+(\frac{2}{\xi _{2}R}+\frac{2}{R^{2}})\frac{\xi _{2}^{2}e^{-R/\xi _{2}}\xi
^{\prime \prime 3}}{\xi _{1}}[2-(R^{2}/\xi ^{\prime \prime 2}+2R/\xi
^{\prime \prime }+2)e^{-R/\xi ^{\prime \prime }}]
\end{equation*}%
\begin{equation*}
-(\frac{2}{\xi _{2}}+\frac{6}{R}+6\frac{\xi _{2}}{R^{2}})\frac{\xi
_{2}^{2}e^{-R/\xi _{2}}\xi ^{\prime \prime 2}}{\xi _{1}}[1-(R/\xi ^{\prime
\prime }+1)e^{-R/\xi ^{\prime \prime }}]
\end{equation*}

\begin{equation*}
+(2+6\frac{\xi _{2}}{R}+6\frac{\xi _{2}^{2}}{R^{2}})\frac{\xi
_{2}^{2}e^{-R/\xi _{2}}\xi ^{\prime \prime }}{\xi _{1}}(1-e^{-R/\xi ^{\prime
\prime }})
\end{equation*}%
\begin{equation*}
+(\frac{2}{R^{2}}-\frac{2}{R\xi _{2}})\frac{\xi _{2}^{2}e^{-R/\xi _{1}}\xi
^{\prime 3}}{\xi _{1}}(\frac{R^{2}}{\xi ^{\prime 2}}+2\frac{R}{\xi ^{\prime }%
}+2)
\end{equation*}%
\begin{equation*}
+(\frac{6\xi _{2}}{R^{2}}+\frac{2}{\xi _{2}}-\frac{6}{R})\frac{\xi
_{2}^{2}e^{-R/\xi _{1}}\xi ^{\prime 2}}{\xi _{1}}(\frac{R}{\xi ^{\prime }}+1)
\end{equation*}%
\begin{equation}
+(2-6\frac{\xi _{2}}{R}+6\frac{\xi _{2}^{2}}{R^{2}})\frac{\xi
_{2}^{2}e^{-R/\xi _{1}}\xi ^{\prime }}{\xi _{1}}\},  \label{vz}
\end{equation}%
where $\xi ^{\prime }$ and $\xi ^{\prime \prime }$ are defined by%
\begin{equation*}
\xi ^{\prime -1}=\xi _{1}^{-1}+\xi _{2}^{-1}\text{ and }\xi ^{\prime \prime
-1}=\xi _{1}^{-1}-\xi _{2}^{-1}.
\end{equation*}%
Eq.(\ref{vz}) displays the exponential decay of velocity matrix elements
with distance $R$ between two localized states. In the variable range
hopping argument\cite{motda}, only the exponential decay of transfer
integral with $R$ is treated. In a process which is first order in transfer
integral, that is not discussed here, one may expect interesting new
features.

Because for each Cartesian component,
\begin{equation}
\langle \phi _{A_{1}}|x_{\alpha }|\phi _{A}\rangle =\frac{i\hbar \langle
\phi _{A_{1}}|v_{\alpha }|\phi _{A}\rangle }{(E_{A}-E_{A_{1}})},\text{ \ }%
\alpha =x,y,z,  \label{rv}
\end{equation}%
from (\ref{env}) and (\ref{vz}), one has%
\begin{equation}
\langle \phi _{A_{1}}|r_{\perp }|\phi _{A}\rangle =0\text{ and }\langle \phi
_{A_{1}}|r_{\parallel }|\phi _{A}\rangle =\frac{i\hbar v_{\parallel
}^{A_{1}A}}{(E_{A}-E_{A_{1}})}\text{.}  \label{1rv}
\end{equation}

\end{document}